\documentclass[aip,
 amsmath,amssymb,
 reprint,
]{revtex4-1}
\usepackage[utf8]{inputenc}
\usepackage{graphicx}
\usepackage{enumitem}
\usepackage{mhchem}
\usepackage{bm}
\usepackage{cases}
\usepackage{upgreek}
\usepackage[usenames, dvipsnames]{color}
\usepackage{hyperref}
\usepackage{braket}
\usepackage{float}
\usepackage[normalem]{ulem}

\usepackage{textgreek}
\usepackage{gensymb}
\usepackage{dcolumn}

\newcommand \rmm[1]  { \textrm{#1} }

\makeatletter
\def\@email#1#2{%
 \endgroup
 \patchcmd{\titleblock@produce}
  {\frontmatter@RRAPformat}
  {\frontmatter@RRAPformat{\produce@RRAP{*#1\href{mailto:#2}{#2}}}\frontmatter@RRAPformat}
  {}{}
}%
\makeatother

\makeatletter
\let\origsection\section
\renewcommand\section{\@ifstar{\starsection}{\nostarsection}}
\newcommand\nostarsection[1]
{\sectionprelude\origsection{#1}\sectionpostlude}
\newcommand\starsection[1]
{\sectionprelude\origsection*{#1}\sectionpostlude}
\newcommand\sectionprelude{%
  \vspace{-6pt}
}
\newcommand\sectionpostlude{%
  \vspace{-10pt}
}
\makeatother
\begin{document}

\preprint{AIP/123-QED}

\title{Hot carrier extraction from 2D~semiconductor photoelectrodes}

\author{Rachelle Austin}
\thanks{These authors contributed equally to this work}
\affiliation{Department of Chemistry, Colorado State University; Fort Collins, CO, USA}
\author{Yusef Farah}
\thanks{These authors contributed equally to this work}
\affiliation{Department of Chemistry, Colorado State University; Fort Collins, CO, USA}
\author{Thomas Sayer}
\thanks{These authors contributed equally to this work}
\affiliation{Department of Chemistry, University of Colorado Boulder; Boulder, CO, USA}
\author{Brad M. Luther}
\affiliation{Department of Chemistry, Colorado State University; Fort Collins, CO, USA}
\author{\\Andr\'{e}s Montoya-Castillo}
\homepage{Andres.MontoyaCastillo@colorado.edu}
\affiliation{Department of Chemistry, University of Colorado Boulder; Boulder, CO, USA} 
\author{Amber Krummel}
\homepage{Amber.Krummel@colostate.edu}
\affiliation{Department of Chemistry, Colorado State University; Fort Collins, CO, USA}
\author{Justin Sambur}
\homepage{Justin.Sambur@colostate.edu}
\affiliation{Department of Chemistry, Colorado State University; Fort Collins, CO, USA}
\affiliation{School of Advanced Materials Discovery, Colorado State University; Fort Collins, CO, USA\looseness=-1}


\date{\today}

\begin{abstract}
Hot carrier-based energy conversion systems could double the efficiency of conventional solar energy technology or drive photochemical reactions that would not be possible using fully thermalized, “cool” carriers, but current strategies require expensive multi-junction architectures. Using an unprecedented combination of photoelectrochemical and \textit{in situ} transient absorption spectroscopy measurements, we demonstrate ultrafast ($<50$~fs) hot exciton and free carrier extraction under applied bias in a proof-of-concept photoelectrochemical solar cell made from earth-abundant and potentially inexpensive monolayer (ML) \ce{MoS2}. Our approach facilitates ultrathin 7~\AA~charge transport distances over 1~cm$^2$ areas by intimately coupling ML-\ce{MoS2} to an electron-selective solid contact and a hole-selective electrolyte contact. Our theoretical investigations of the spatial distribution of exciton states suggest greater electronic coupling between hot exciton states located on peripheral S atoms and neighboring contacts likely facilitates ultrafast charge transfer. Our work delineates future 2D~semiconductor design strategies for practical implementation in ultrathin photovoltaic and solar fuels applications.
\end{abstract}

\maketitle

\section{Introduction}

All semiconductors absorb photon energies greater than their bandgap, temporarily creating hot carriers with excess energy. In a conventional solar cell material like Si, 40\% of the absorbed solar energy is lost as heat because hot carriers rapidly cool in $<100$~fs.\cite{Nozik2003} This ultrafast hot-carrier cooling process prevents current solar cell technology from reaching theoretical efficiency limits.\cite{Shockley2004}

A longstanding challenge in the field is to develop materials and selective charge-extraction contacts that efficiently collect hot carriers before they cool.\cite{Nozik2018} The first experimental demonstration of hot-carrier extraction involved the transfer of photogenerated hot carriers from a bulk InP crystal to p-nitrobenzonitrile molecules in an electrochemical cell.\cite{Cooper1998} Unfortunately, a significant fraction of hot-electrons thermalized in the InP bulk. Recent efforts in solid-state photovoltaics have focused on enhancing hot carrier populations in the semiconductor absorber and using charge carrier-selective contacts to preferentially extract hot electron and hole populations (e.g., \ce{In_{0.78}Ga_{0.22}As_{0.81}P_{0.19}} quantum well surrounded by thick \ce{In_{0.8}Ga_{0.2}As_{0.435}P_{0.565}} barriers contacted by n- and p-doped InP contact layers)\cite{Nguyen2018}. Unfortunately, such multi-layer structures require expensive materials and growth methods, especially for the critical charge-selective contacts.

\begin{figure*}[!t]
\vspace{-18pt}
\begin{center}
    \resizebox{.7\textwidth}{!}{\includegraphics{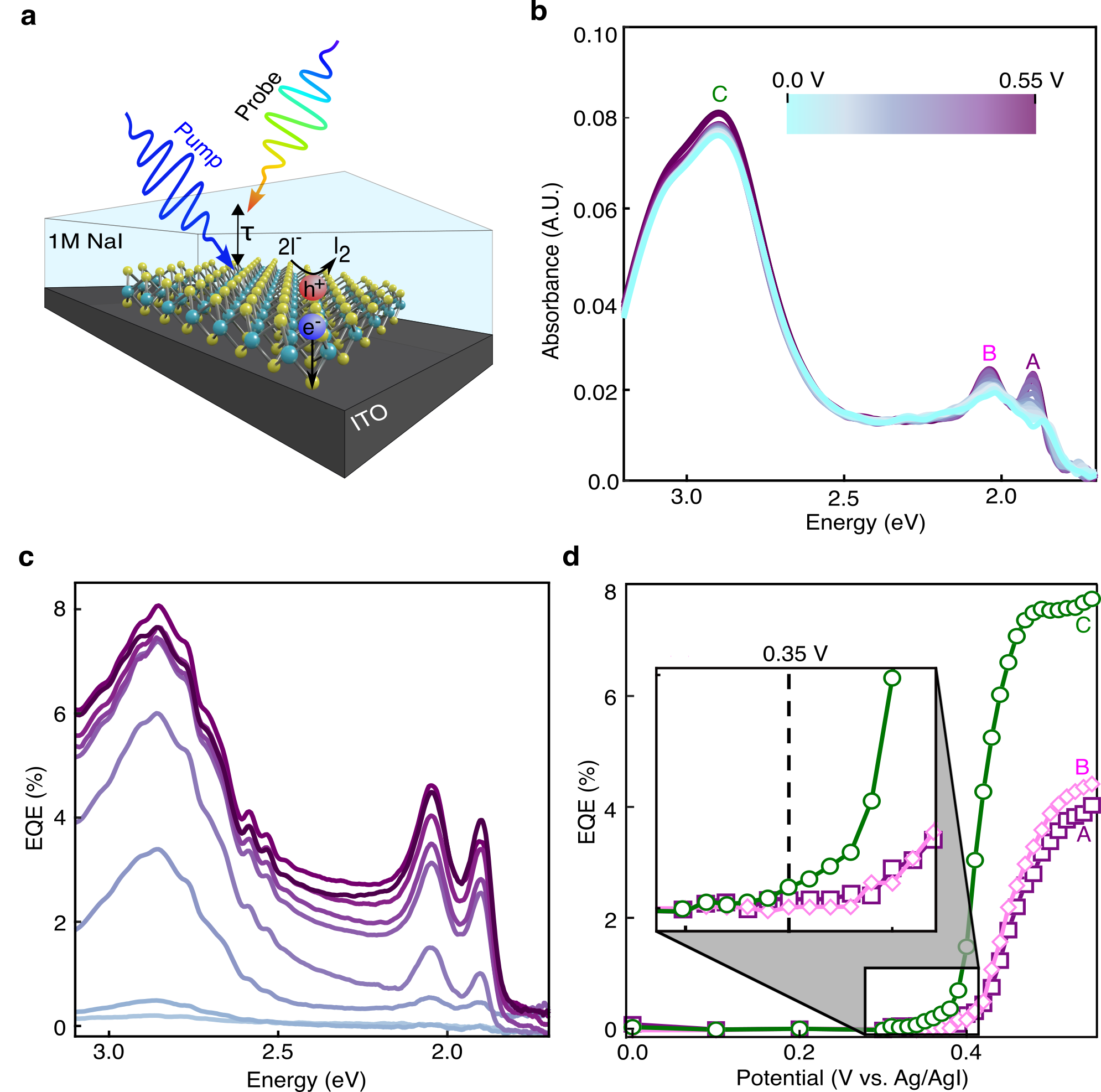}}
\end{center}
\vspace{-10pt}
\caption{\label{fig:1} Optoelectronic properties of the monolayer \ce{MoS2} photoelectrochemical cell. \textbf{(a)} Cartoon illustration of the three-electrode photoelectrochemical cell. The solid blue and rainbow arrows indicate pump and probe pulses for TA measurements. Pt counter and Ag/AgI reference electrodes are omitted for clarity. \textbf{(b)} Absorbance spectra in $0.025~$V increments from $0.000$~V to $0.550~$V.  \textbf{(c)} EQE spectra versus applied potential from $0.35~$V to $0.55~$V. $EQE(\lambda)=qi/I_0(\lambda)$, where $q$ is the electronic charge (in units of~C), $i$ is the photocurrent (in units of~A), and $I_0$ is the monochromatic light power (in units of~$s^{–1}$). \textbf{(d)} Monochromatic $i$-$E$ measurements for resonant A-, B-, and C-exciton excitation (i.e., $650~$nm, $605~$nm, and $405~$nm, respectively).}
\vspace{-12pt}
\end{figure*}

Nanostructured materials possess unique photophysical and structural properties that could make hot-carrier-based energy conversion systems both efficient and inexpensive, but this remains a formidable challenge.\cite{Nozik2021a} For example, while hot carrier extraction in graphene optoelectronics is possible,\cite{Gabor2011} graphene-based photovoltaics exhibit low quantum efficiency and small photovoltages.\cite{Sun2012a, Xu2010b} Plasmonic metal nanostructure solar energy conversion systems suffer from low power conversion efficiency due to fast hot carrier cooling.\cite{Zhang2016d} While hot carrier extraction at model interfaces made of solution-processed organic–inorganic lead halide perovskite\cite{Li2017a} and lead chalcogenide nanocrystals\cite{Tisdale2010} has been demonstrated, electrical measurements of hot-carrier effects in a working solar energy conversion system are scant. Indeed, recent ultrafast X-ray measurements of lead halide perovskites highlight the need for concurrent electrical measurements because interpretation and quantification of hot electron and hole temperatures can be difficult using optical measurements alone.\cite{Verkamp2021}

In this work, we investigate the intriguing possibility of using inexpensive, earth-abundant, and potentially scalable\cite{Wadia2009} transition metal dichalcogenides (TMDs) such as monolayer (ML) \ce{MoS2} for hot carrier extraction using a proof-of-concept photoelectrochemical solar cell. The bulk \ce{MoS2 | I^-, I3^-|Pt} photoelectrochemical cell is a stable $>14$\%-efficient solar cell\cite{Kline1981} that is limited by a small $0.6$~V photovoltage, which hot carrier collection could potentially circumvent. ML TMDs are exciting absorber materials because high energy photons generate hot excitons, often called C-excitons, with $>100$~ps lifetimes.\cite{Wang2017e} These long and tunable\cite{Rose2022} lifetimes exceed ultrafast photocurrent response in optoelectronic devices,\cite{Vogt2020} which suggests hot carriers could be contributing to current in the device. Optical signatures of hot carrier transfer from \ce{MoS2} to graphene\cite{Wang2017e} and gold\cite{GrubisicCabo2015} also suggest hot carrier transfer could outpace cooling, but electrical signatures of hot carrier transfer have remained elusive. Electrical measurements of solid-state ML TMD devices typically require edge-on contacts for charge extraction, meaning charge carriers travel micron-long distances to the charge-collecting interfaces. The long transport distances promote cooling and recombination, which likely explains why hot carrier-induced currents have not been reported in any 2D TMD-based solar energy conversion device. The advantage of employing TMD absorbers in a photoelectrochemical cell is that photogenerated charge carriers need only travel across three atoms ($0.7$~nm) to reach the electron-selective indium tin oxide (ITO) substrate and hole-collecting redox electrolyte (Fig.~\ref{fig:1}a). The long lifetime of C-excitons combined with short charge-transfer distances and intimate charge carrier-selective contacts thus raises the exciting possibility of extracting hot carriers in the model ML-\ce{MoS2 | I^-, I3^-|Pt} solar cell.

Using an unprecedented combination of photoelectrochemical and \textit{in situ} time-resolved spectroscopic techniques, here we show hot carrier extraction outcompetes exciton formation and relaxation in a working photoelectrochemical cell. We demonstrate that hot carriers can be extracted to generate photocurrent before cooling to the band-edge and develop a picture of the photocurrent generation mechanism in ML-\ce{MoS2} photoelectrodes.


\section{Optical and Photoelectrochemical Characterization of the ML-M\texorpdfstring{\MakeLowercase{o}}{o}S$_2$ photoelectrode}

Here we test our hypothesis that one should be able to preferentially extract hot carriers from the C-exciton manifold in ML-\ce{MoS2} using the electron- and hole-selective contacts in a photoelectrochemical cell because the three atom-thick transport distance minimizes the electron and hole transport times relative to the cooling times. Photoluminescence and Raman spectroscopy confirmed the chemical vapor deposition (CVD) growth method produced ML-\ce{MoS2} (Appendix~\ref{app:chara}, Fig.~\ref{fig:si1}a,b). To probe the hot-carrier generation, recombination, and extraction processes, we constructed a transparent photoelectrochemical cell that enables \textit{in situ} ultrafast spectroscopy measurements to measure relative exciton populations after photo-excitation (Fig.~\ref{fig:1}a and Appendix~\ref{app:constr}). The ML-\ce{MoS2}-coated ITO electrode serves as the working electrode in a three-electrode microfluidic electrochemical cell containing 1~M~\ce{NaI} electrolyte. Fig.~\ref{fig:1}a schematically shows that, under illumination and applied positive potentials, photogenerated holes move to the semiconductor/liquid interface and oxidize \ce{I^-} to \ce{I2} while photogenerated electrons transfer to the ITO substrate, generating net anodic current flow through the cell.

\begin{figure}[!b]
\vspace{-24pt}
\begin{center}
    \resizebox{.4\textwidth}{!}{\includegraphics{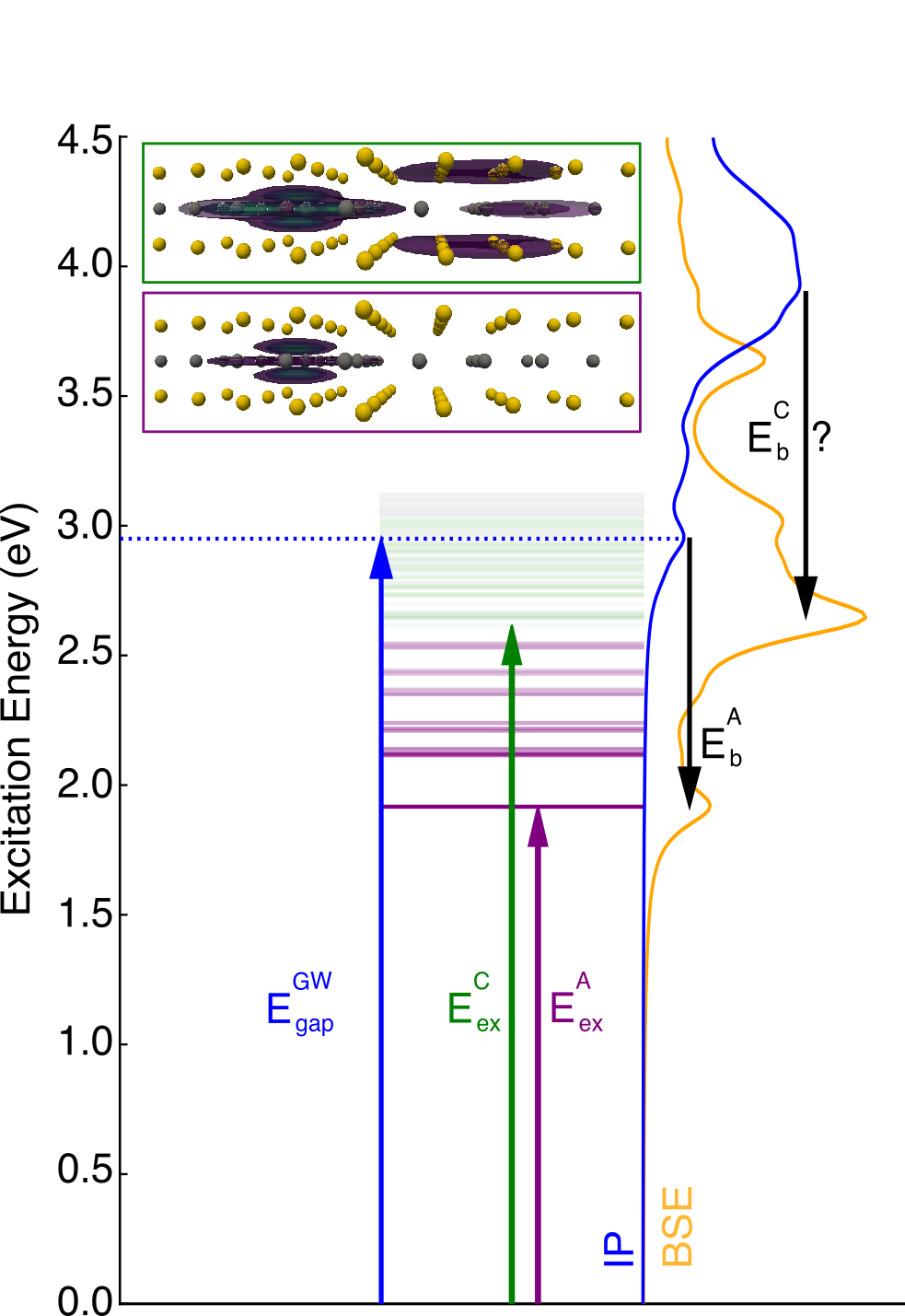}}
\end{center}
\vspace{-4pt}
\caption{\label{fig:2} \textbf{Center}: Eigenspectrum of unsupported ML-\ce{MoS2} in vacuum calculated from BSE. The purple and green lines represent the A-exciton `hydrogenic series’ around the K~point and the C-exciton states in the band nesting region between K and $\Gamma$, respectively. The transparency of each energy level is set to the oscillator strength of the transition, normalized by the A-1s transition (labeled $E_\rmm{ex}^A$), which is the strongest. $E_\rmm{ex}^C$ is the energy of the first C-exciton state. The dotted blue line represents the fundamental electronic band gap at $2.9$~eV. \textbf{Top}: The hole-averaged isosurface plots for the electron density in the aforementioned A-exciton (below, purple box) and C-exciton (above, green box) states. \textbf{Right}: The absorption spectra as calculated for the BSE states pictured (orange) and the underlying single, independent particle states (blue). The energy difference between peaks represents the A- and C-exciton binding energies ($E_\rmm{b}^A$ and $E_\rmm{b}^C$).}
\end{figure}

To assess which excitonic transitions contribute to current flow in the ML-\ce{MoS2} photoelectrode, we simultaneously measured optical absorbance and photocurrent signals under working photoelectrochemical conditions (Appendix~\ref{app:echemMeasure}). Fig.~\ref{fig:1}b shows absorbance spectra of ML-\ce{MoS2} as a function of the applied potential ($E$, referenced to the Ag/AgI electrode). All spectra feature three peaks corresponding to the band-edge A-~($1.99$~eV) and B-~($2.01~$eV) excitons, and the higher energy C-exciton ($2.98$~eV). The A- and B-exciton peak intensities increase and blue shift with increasing positive potential, while the high energy C-exciton increases slightly with positive bias and does not shift with potential. The observed potential-dependent absorbance changes (Fig.~\ref{fig:1}b) have important consequences for interpreting the relative populations of excitons gleaned from ultrafast TA measurements, as will be discussed below.

Photocurrent measurements performed concurrently with the absorbance measurements revealed at which potentials the different excitons dissociate and contribute to current flow in the cell. Figure~\ref{fig:1}c shows potential dependent external quantum efficiency (EQE) spectra, where $EQE(\lambda)=qi/I_0(\lambda)$ and $i$ is the photocurrent, $q$ is the elementary charge, and $I_0$ is the light power. At positive bias (e.g., $E > 0.5$~V), the EQE spectrum mimics the absorbance spectrum (solid purple line in Fig.~\ref{fig:1}c), indicating the applied potential generates a sufficiently strong interfacial electric field to effectively dissociate all three excitonic species. Close examination of the potential-dependent spectra reveals subtle differences between the photocurrent onset potential for the A-, B- and C-excitons that could be further distinguished in monochromatic $i$-$E$ curve measurements (Fig.~\ref{fig:1}d). Interestingly, photocurrent generation onsets first (i.e., less positive potentials) for the C-exciton and the slope of the C-exciton $i$-$E$ curve is significantly steeper than that of the A/B-excitons. The lower onset potential means that, under conditions of equivalent interfacial electric field strength, photo-excited C-excitons require less driving force to dissociate and contribute to current flow in the cell compared to the lower energy A/B-excitons.\cite{Klots2014} Since we did not observe photocurrent upon directly exciting the A/B-excitons at the same applied potential as that of the C-exciton (specifically $E=0.35$~V in Fig.~\ref{fig:1}d), our monochromatic $i$-$E$ data strongly suggests that C-excitons are extracted before they cool to the band edge and form low energy A/B-excitons.

\section{Nature of excitonic states and their implication for charge transfer}

Why is C-exciton extraction more efficient than low energy A/B-exciton extraction in our ML-\ce{MoS2} photoelectrode? To shed light on this critical question, we turned to a theoretical treatment of the spatial distribution of the A/B- versus C-excitons in ML-\ce{MoS2}. The goal of this theoretical investigation is to qualitatively compare the magnitude of electronic coupling between the different exciton states in the charge donor (\ce{MoS2}) and the electron/hole acceptors (ITO substrate/iodide anions), which ultimately allows us to rationalize why charge transfer from the C-exciton is more efficient than A/B-excitons. 

We first analyzed the excitonic wavefunctions obtained by solving the Bethe-Saltpeter equation (BSE) for unsupported ML-\ce{MoS2} in vacuum (see Appendix~\ref{app:bse}). This approach accounts for excitonic effects by explicitly treating electron-hole correlation. Figure~\ref{fig:2}-center shows the resulting eigenspectrum of pristine ML-\ce{MoS2}, where the bound eigenstates arising from a BSE calculation are linear combinations of independent-particle states at different points in quasi-momentum (k-) space (see Appendix Fig.~\ref{fig:si8} for visualisation of the composition of three of the states). The purple lines are the lowest energy excitonic states around the K point. The lowest energy transition and optical bandgap of ML-\ce{MoS2} is the A-exciton ($E_\rmm{ex}^A$) while the higher energy states are the A-exciton ‘hydrogenic series', consistent with literature.\cite{Qiu2013, Molina-Sanchez2013} The green lines represent the levels in the band nesting region between K~and~$\Gamma$, where the lowest energy transition is the C-exciton $E_\rmm{ex}^C$. The B-exciton states do not appear in Fig.~\ref{fig:2}-center because our calculations do not include spin-orbit coupling effects.

\begin{figure*}[!t]
\vspace{-14pt}
\begin{center}
    \resizebox{.7\textwidth}{!}{\includegraphics{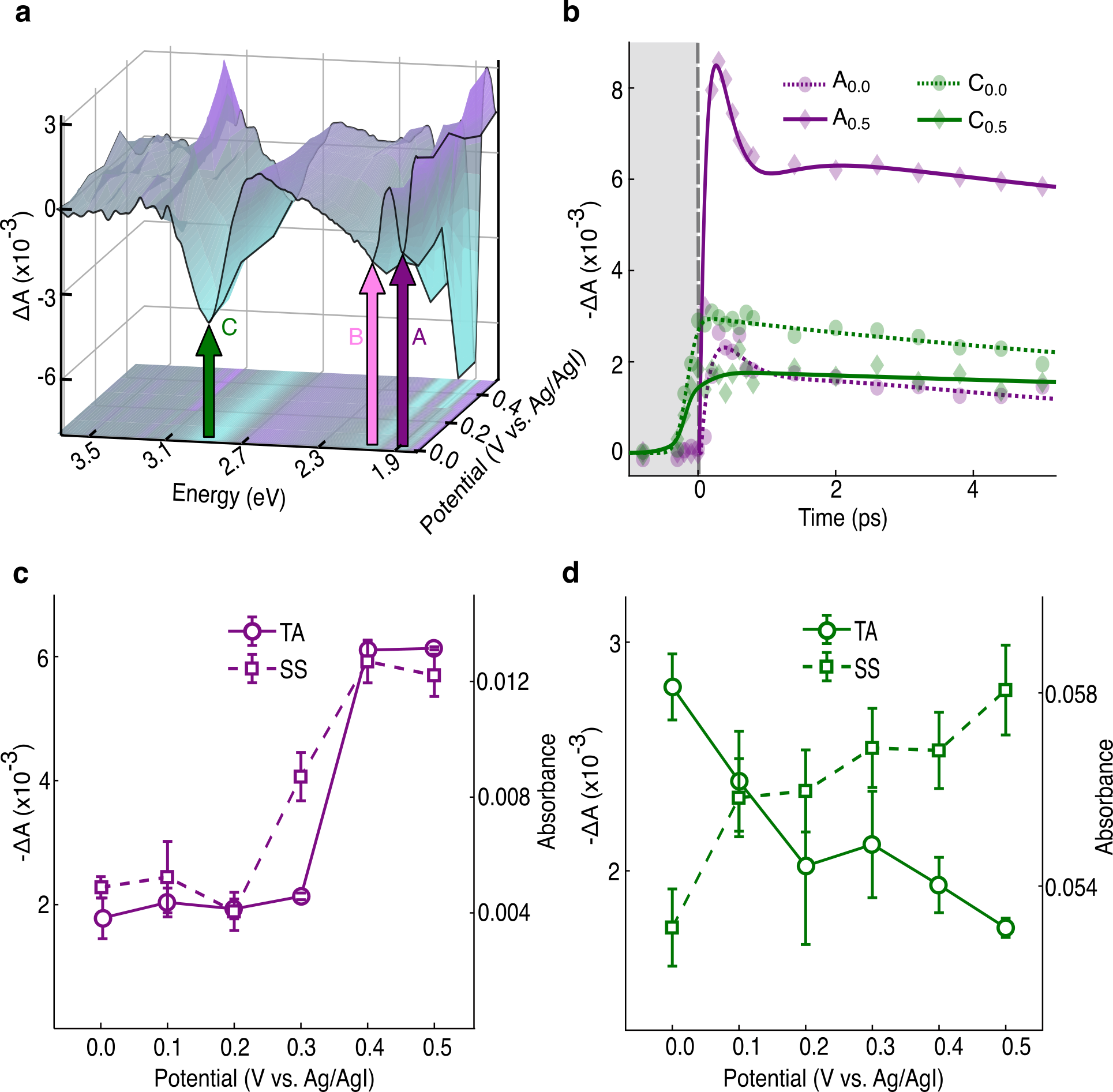}}
\end{center}
\vspace{-10pt}
\caption{\label{fig:3} \textbf{(a)} Potential-dependent TA spectra at $\tau = 1$~ps. \textbf{(b)} Temporal trace of the A-exciton (purple) and C-exciton (green) bleach intensity for $0.0$~V (filled circles) and $0.5$~V (filled diamonds). Solid and dashed lines are fits to Eqs.~\ref{eq:1} and \ref{eq:2} in Appendix~\ref{app:decay}. \textbf{(c,d)} Steady-state absorbance (squares) and TA bleach intensity (circles) for the \textbf{(c)} A- and \textbf{(d)} C-excitons.}
\vspace{-6pt}
\end{figure*}

\vfill

We now consider the comparative ease of extracting electrons from the photo-excited C-exciton in contrast to A/B-excitons from a classic Marcus-type charge transfer\cite{Marcus1993, Chidsey1991a} pathway comparing the relative magnitude of the electronic coupling associated with electron (hole) transfer from the C- vs A/B-excitons in \ce{MoS2} to ITO (iodide).\footnote{Marcus-type arguments have been previously employed to analyze charge transfer characteristics between TMDCs and quantum dots or molecular species.} When the rate of electron (hole) transfer to the ITO substrate (iodide) is in the small electronic coupling regime, then the rate of charge transfer is proportional to $|V|^2$, the absolute value of the electronic coupling that connects the donor and acceptor states. The A- and C-excitons, formed from states in different regions of the band structure, have different spatial distributions within the \ce{MoS2} layer. Figure~\ref{fig:2}-top illustrates that A-excitons are localized on interior Mo~atoms whereas C-excitons have significant electron density on the peripheral S~atoms, consistent with the atomic orbital composition of these k-space regions in previous DFT calculations.\cite{Fang2015, Kadantsev2012, Kormanyos2015} Since $V$~decays exponentially with the distance between the donor and acceptor, one would expect that the C-excitons’ closer approach to the ITO/aqueous interfaces results in a faster charge transfer rate. 
\vfill
To make this a quantitative Marcus-type analysis, one would additionally require the construction of charge localized states, the calculation of their energies and couplings, the calculation of exciton/charge phonon coupling, and its modulation by the disorder of the solution. Methods for these calculations are still under development\cite{Migliore2012, Ghosh2016c, Giustino2017, Zhou2021, Chen2020d} and, therefore, beyond the scope of this study. Nevertheless, our results and calculations provide unique insight into the origin of distinct photocurrent efficiencies for the A/B- and C-excitons in a ML-\ce{MoS2} photoelectrode. Furthermore, since the hole density of the C-exciton is localized on the interior Mo atoms and less electronically coupled to the iodide acceptors, one might hypothesize that the redox mechanism is a sequential one of electron transfer to ITO preceding hole transfer to the iodide acceptor in solution. We now turn to ultrafast optical measurements to probe the mechanism of photocurrent generation in greater detail.

\vfill\vfill\vfill\pagebreak
\section{Exciton Population Dynamics via Ultrafast Transient Absorption Spectroscopy}

We leveraged ultrafast pump-probe transient absorption (TA) spectroscopy to assess the feasibility and elucidate the mechanism of hot carrier extraction in ML-\ce{MoS2} photoelectrodes. \textit{in situ} TA spectroscopy allowed us to measure relative changes in exciton populations under working photoelectrochemical conditions. In this experiment, we employ a transform limited, $<50$~fs, $3.1$~eV pump pulse corresponding to the high energy side of the C-exciton to generate a distribution of excited states, including hot free carriers in continuum states of ML-\ce{MoS2}. Experiments were performed with a pump fluence of 75~µJ/cm$^2$, where the exciton bleach intensities are linear with pump fluence (Appendix Fig.~\ref{fig:si2}), in agreement with literature.\cite{Wang2017e} A white light continuum probe pulse spanning 3.1~eV--1.65~eV was used to measure the absorbance changes of the sample as a function of pump-probe delay time ($\tau$) (see Appendix~\ref{app:taMeasure}). 

A potential-dependent anodic current flows through the cell upon photoexcitation of the sample with the pump pulse (Appendix Fig.~\ref{fig:si3}), further confirming that the TA experiment reports on exciton populations under working photoelectrochemical conditions. Figure~\ref{fig:3}a shows TA spectra at $\tau = 1$~ps as a function of applied potential. The $\tau = 1$~ps condition yields spectra whose excitonic features are not dominated by the influence of hot electrons or free carriers and, therefore, this delay time provides a clear picture of the relative changes in exciton population versus applied potential.\cite{Cunningham2017} The TA signal is given by $\Delta A(\tau)=A_\rmm{pumped}(\tau)-A_\rmm{unpumped}$, where $A_\rmm{pumped}(\tau)$ and $A_\rmm{unpumped}$ represent the absorbance values of the pumped and unpumped samples, respectively.  Figure~\ref{fig:3}a shows bleach features (negative $\Delta A(\tau)$) at the exciton peak energies due to exciton formation and some positive  features due to pump induced lineshape narrowing and excited state absorption. Unlike bulk semiconductors, $A_\rmm{unpumped}$ changes significantly with applied potential (Fig.~\ref{fig:1}b) for ML-\ce{MoS2} and must be considered when interpreting relative exciton populations from \textit{in situ} TA data.

Before discussing changes in the relative exciton populations, we first examine the exciton formation and depletion processes in the ML-\ce{MoS2} photoelectrode. Figure~\ref{fig:3}b shows A- and C-exciton TA decay dynamics at $+0.5$~V and $0.0$~V, corresponding to conditions of significant anodic photocurrent and no photocurrent, respectively. The B-exciton behavior closely follows that of the A-exciton (Appendix Fig.~\ref{fig:si4}) and is omitted from Fig.~\ref{fig:3}b for visual clarity. The temporal behavior of the C-exciton and the band edge A/B-excitons display three key differences (see Fig.~\ref{fig:3}b, Appendix Figs.~\ref{fig:si4},\ref{fig:si5}, and Appendix Table~\ref{tab:1}). First, the C-exciton bleach intensity reaches a maximum value almost instantaneously (i.e., $\tau = 0$~ps), indicating C-excitons form on a faster timescale than our instrument response ($<50$~fs), which aligns with previous work\cite{Trovatello2020} and is expected since the pump pulse is tuned to the high energy side of the C-exciton spectrum. Second, the A-exciton bleach reaches a maximum value at longer times ($\tau = 200$~fs, Appendix Table~\ref{tab:1}). During this time, we observe no decay of the C-exciton bleach, indicating that C-excitons do not cool to A-excitons on the $200$~fs timescale; if C-excitons decayed to A-excitons over this timescale, one would expect to observe a fast C-exciton decay component commensurate with the A-exciton rise time. Instead, after C-exciton formation, the fast component for C-exciton decay occurs within the first tens of picoseconds (Appendix Table~\ref{tab:1}]), consistent with intervalley relaxation mechanisms.\cite{Wang2017e, Rose2022} Third, the C-exciton bleach magnitude at $\tau= 0$~ps is greater (i.e., more negative) at $0.0$~V than $+0.5$~V, but the opposite trend occurs for the A-exciton. This third observation is critical for interpreting how the exciton populations change as a function of applied potential and will be discussed below.

We now exploit steady-state absorbance and ultrafast TA measurements to disentangle the relative exciton population changes. Recalling the TA signal, $\Delta A(\tau)=A_\rmm{pumped}(\tau)-A_\rmm{unpumped}$, and recognizing that steady-state absorbance measurements represent $A_\rmm{unpumped}$, we extract relative changes in the exciton populations from $A_\rmm{pumped}$, where a larger magnitude indicates more exciton population. Figure~\ref{fig:3}c,d compares steady-state absorbance and TA bleach intensities of the A- and C-excitons as a function of applied potential. For the A-exciton (Fig.~\ref{fig:3}c), the TA and steady-state data closely follow one another, suggesting that the increase in $\Delta A(\tau)$ can be attributed to $A_\rmm{unpumped}$ instead of $A_\rmm{pumped}$. In contrast to the A-exciton, $A_\rmm{unpumped}$ for the C-exciton slightly increases with $+E$ (Fig.~\ref{fig:1}b), but $\Delta A(\tau)$ significantly decreases for the C-exciton (Fig.~\ref{fig:3}d), suggesting that fewer C-excitons contribute to the TA signal under these conditions.


Two scenarios could explain why C-excitons contribute less to the TA signal with $+E$. First, bound C-excitons could dissociate, producing photocurrent in the external circuit via the Marcus-type scheme introduced above. This scenario requires that photocurrent generation occurs faster than TA signal generation. The second possibility is that a portion of the hot, free electrons created by the pump pulse in the continuum levels of the conduction band do not cool to form C-excitons. Instead, they inject directly into the ITO electrode before C-exciton formation and generate photocurrent. This latter possibility offers exciting opportunities to achieve solar energy conversion efficiencies beyond the detailed balance limit. 
Theoretically, the latter mechanism of hot electron injection into the ITO is compatible with a scenario where bound C-excitons and continuum unbound carriers interconvert through isoenergetic scattering processes. Our calculations indicate that the binding energy of C-excitons is on the order of that of the A-exciton, making them degenerate with unbound continuum states elsewhere in the Brillouin zone, an observation consistent with previous calculations.\cite{Klots2014, Aleithan2016} Since these bound and unbound species are degenerate, one may imagine highly efficient intra- and inter-band scattering, known to occur on timescales of tens to hundreds of femtoseconds,\cite{Gao2021} that would establish a dynamic equilibrium between C-excitons and free carriers. These unbound species could then be extracted as hot carriers by a moderate applied field or subsequently relax to the band-edge, where they would only contribute to the current at sufficiently high applied fields. In contrast, A-excitons (and their hydrogenic series) are necessarily mid-gap, and so cannot be degenerate with unbound carriers. This means that to extract an electron or hole from these species, one will always incur the additional cost of overcoming their binding energy, presumably explaining the additional driving force (more positive potential) required for A/B-exciton photocurrent generation in Fig.~\ref{fig:1}d-inset. Currently, \textit{ab~initio} methods are being developed to calculate the scattering matrix elements that would make an in-depth analysis of such an argument possible, albeit primarily on charge-neutral systems.\cite{Efimkin2021, Ataei2021} 

While our theoretical treatment is presently useful in delineating the possibilities of how hot carrier extraction can occur in ML-\ce{MoS2}, we emphasize that these calculations are only valid for unsupported \ce{MoS2} in vacuum. Indeed, it is challenging to treat the full system accurately within the BSE, since one must include the role of the ITO substrate and electrolyte screening, the finite carrier population (as a function of potential), phonon coupling, and excitonic band-dispersion (in the nesting region in particular) into an already costly computation. The aforementioned factors are in fact major areas of current research\cite{Chang2019b, Cho2022, Antonius2022} and the results presented in this paper strongly motivate their continued study. Beyond the fundamental challenge of an affordable, unified electronic structure of this material, a full theoretical treatment will need to address the statistical, electrochemical aspects of the device, meaning this is only the first step in the construction of a predictive multiscale model of photoelectrochemical activity in ultrathin TMDs.

\vspace{-8pt}
\section{Conclusion}

Steady state and ultrafast spectro-electrochemical experiments support an ultrafast, hot carrier extraction mechanism in a ML-\ce{MoS2} photoelectrochemical cell. We employed theoretical treatments to construct a photophysical model of photocurrent generation and hot carrier extraction. In our working model, the extraction of hot carriers following high energy photo-excitation can be accomplished before the formation of excitonic species, leading to the high photocurrent yields in our proof of concept photoelectrochemical solar cell architecture. Calculation of excitonic states suggested two possible mechanisms: one arising from a classic Marcus picture of charge transfer that implies that the delocalized orbital composition for high energy C-excitons may facilitate efficient and ultrafast carrier extraction within the photoelectrochemical cell, and the other invoking a rapid equilibrium between bound C-excitons and isoenergetic free carriers that can be extracted more easily than their bound counterparts. Importantly, our experimental results present a grand challenge to existing theory and motivate the continued improvement of many-body theories in 2D materials and their heterostructures.

The mechanisms we suggest delineate the development of 2D~semiconductors for practical implementation in ultrathin photovoltaic and solar fuels applications. A lightweight ML 2D~semiconductor photovoltaic device can achieve $70~$kW/kg of specific power,\cite{Islam2022} which can be further improved with top and bottom n- and p-type electrical contacts whose energy levels are optimized for hot carrier extraction\cite{Wang2022a} and the engineering of 2D~semiconductor/substrate architectures for maximum light absorption.\cite{Green2003}  Our work also highlights opportunities to develop large-area solid-state systems that intimately couple the ultrathin semiconductor to charge carrier-selective contacts without short-circuiting the device. For photocatalysis and solar fuels research, we envision Janus-type 2D~semiconductor absorbers where a reductive co-catalyst for \ce{H2} production or \ce{CO2} reduction is attached to one layer of S atoms and an oxidative co-catalyst is positioned on the opposing S~layer; the redox potentials of the co-catalysts can be tuned relative to the excited state electron and hole levels in the 2D~semiconductor. There are rich opportunities to study how the energy level alignment between redox species in the liquid electrolyte and hot electron/hole states of ML-\ce{MoS2} affect hot carrier extraction rates and yields.

\section{Acknowledgements}
R.A.~acknowledges support with sample preparation from Michael Van Erdewyk. This research was supported by the U.S.~Department of Energy, Office of Science, Office of Basic Energy Sciences, under Award DE–SC0021189 (J.B.S., R.A.), and under Award DE-SC0016137 (A.T.K., Y.R.F.). A.M.C.~acknowledges the start-up funds from the University of Colorado, Boulder. A.M.C. and T.S.~thank Sivan Refaely-Abramson for helpful discussions and feedback on the manuscript. T.S.~also thanks the speakers and organizers of the ICTP~2022 hybrid meeting ``Ab-initio Many-Body Methods and Simulations with the Yambo Code'' for their time.
\vfill
\pagebreak
\vspace{-14pt}
\appendix*
\section{Materials and Methods}
\subsection{Sample Characterization}\label{app:chara}
\vspace{-10pt}

\begin{figure}[!h]
\vspace{+32pt}
\begin{center}
    \resizebox{.45\textwidth}{!}{\includegraphics{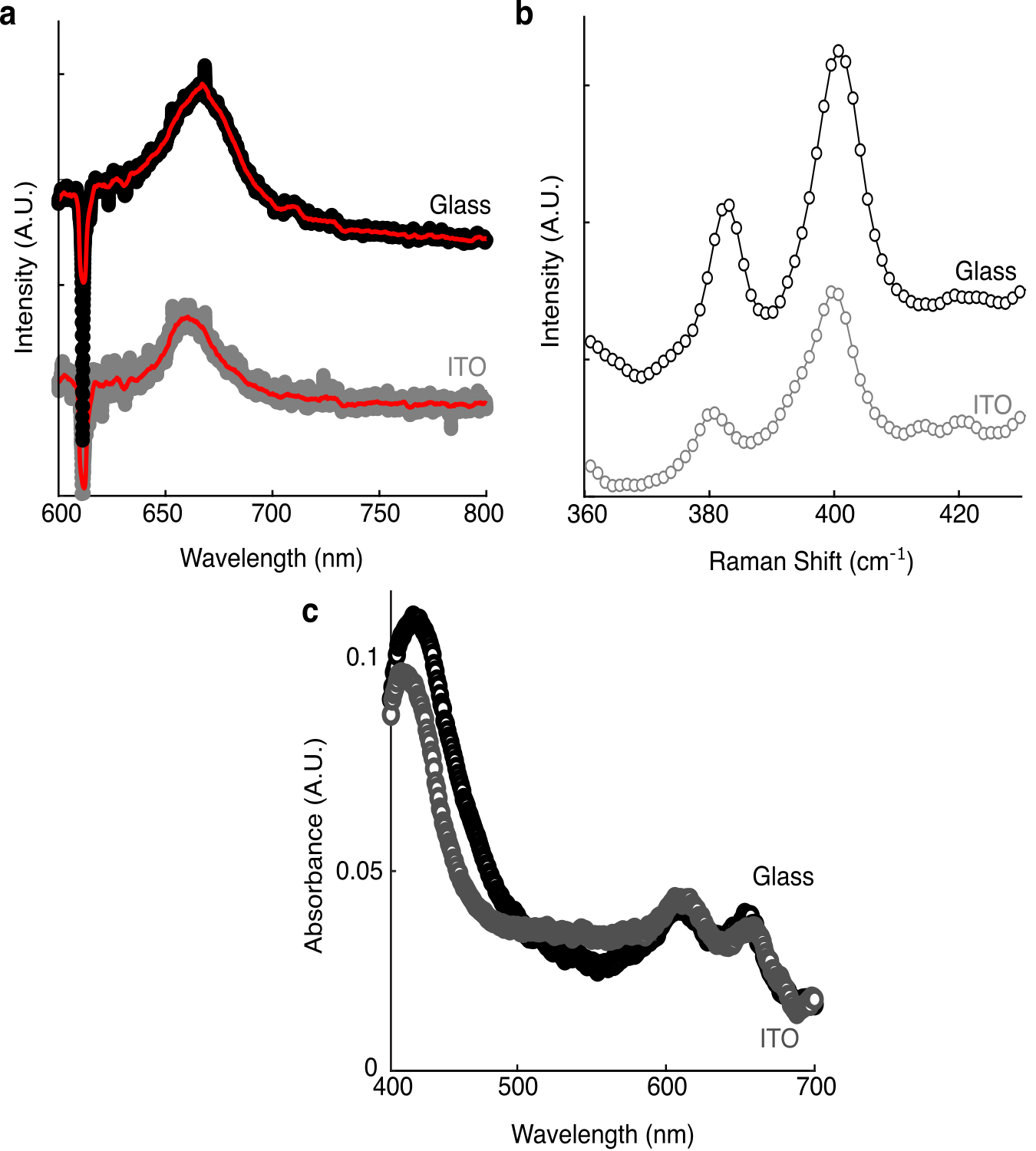}}
\end{center}
\vspace{-14pt}
\caption{\label{fig:si1} Characterization of \ce{MoS2} thin film samples. \textbf{(a)} Photoluminescence spectra of monolayer \ce{MoS2} on ITO in air. The spectrum was acquired using 8.1~MW/cm2 532~nm laser excitation. The red lines represent a 5-point smooth trend line. The PL peak at 660~nm (1.88~eV) is consistent with ML-\ce{MoS2}.\cite{Lee2010} The PL intensity is quenched on the conducting ITO substrate, likely due to additional recombination pathways or charge transfer on ITO.\cite{Lee2014a} \textbf{(b)} Raman spectra of monolayer \ce{MoS2} on ITO. The peaks at 380~cm$^{–1}$ and 400~cm$^{–1}$ are the E$_{2\rmm{g}}^1$ and $A_{1\rmm{g}}$ modes of ML-\ce{MoS2}, respectively, and the 20~cm$^{–1}$ peak-splitting is also consistent with ML-\ce{MoS2}.\cite{Lee2010} \textbf{(c)} Optical Density (OD) spectra of the ML-\ce{MoS2} sample.}
\vspace{-12pt}
\end{figure}

\begin{figure}[!b]
\vspace{-8pt}
\begin{center}
    \resizebox{.3\textwidth}{!}{\includegraphics{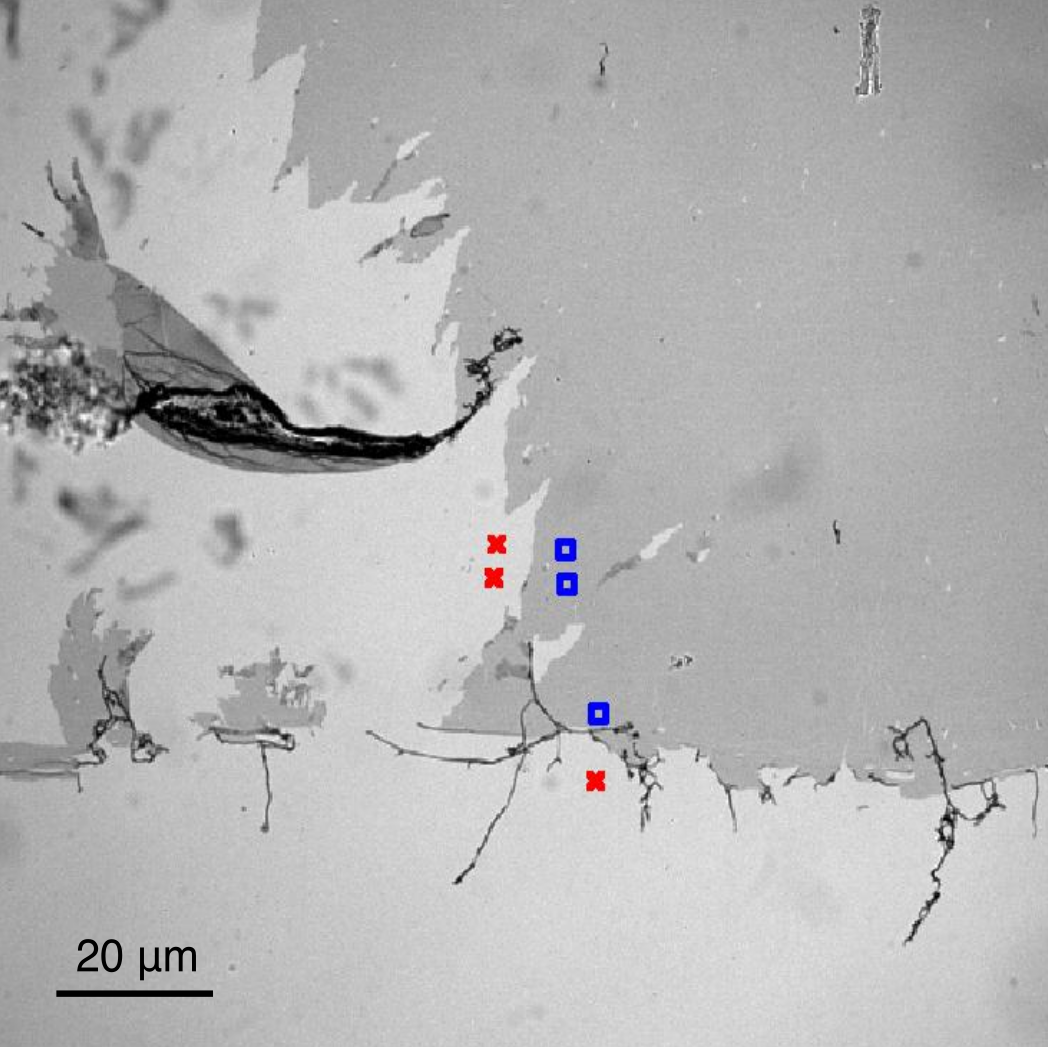}}
\end{center}
\vspace{-14pt}
\caption{\label{fig:si6} Hyperspectral absorbance imaging of the ML-\ce{MoS2} electrode. Representative bright field transmission image acquired under monochromatic illumination. The red and blue regions represent the light transmitted through the ITO substrate and ML-\ce{MoS2} material, respectively. Note, these data are acquired at the film edge to obtain $I_0$ data. }
\end{figure}

Monolayer \ce{MoS2} films (1~cm x 1~cm, 6Carbon Technology, Shenzhen, China) were synthesized via chemical vapor deposition (CVD) on sapphire substrates and mechanically transferred to ITO-coated glass slides using a polymethyl methacrylate stamp and stored in nitrogen-purged glovebox until use. Raman and photoluminescence (PL) micro-spectroscopy experiments confirmed the commercial samples were monolayer \ce{MoS2} (Appendix Fig.~\ref{fig:si1}). The PL and Raman spectra were measured on an inverted Olympus IX73 optical microscope by directing an Ondax 532~nm laser excitation source through a 100×~NA0.95 air objective (Olympus PlanFL N100X) onto the sample. Both signals were collected in a backscatter geometry, filtered by the Ondax~532~nm THz Raman system to remove the fundamental excitation light, passed through a Horiba iHR 550~spectrometer, and then detected by a Synapse charge-coupled device detector. Spatially resolved \textit{in situ} UV-Vis measurements were performed using the procedures described in our previous publications.\cite{Wang2019f} The sample was excited via monochromatic light from a Horiba OBB Tunable PowerArc Illuminator. A 20× microscope objective collected the light transmitted through the sample and a Photometrics Prime 95B~back illuminated complementary metal–oxide–semiconductor (CMOS) camera acquired widefield images of the sample at different monochromatic excitation wavelengths. Absorbance spectra were calculated from the hyperspectral imaging data by spatially selecting the light transmitted through the sample ($I$) and the ITO substrate ($I_0$), where $A(\lambda) = -\log_{10}(I(\lambda)/I_0(\lambda))$. Appendix Fig.~\ref{fig:si6} shows a representative image of the sample as well as $I$ and $I_0$ regions used for absorbance spectra calculations. Note, absorbance data was acquired at the film edge to obtain $I_0$ measurements. TA data was acquired in the film interior where the sample morphology is $>90$\% pure ML-\ce{MoS2}.\cite{Wang2019f}

\vspace{-22pt}
\subsection{Photoelectrochemical Flow Cell Construction}\label{app:constr}
\vspace{-12pt}
\ce{MoS2}-coated ITO substrates were constructed into three-electrode microfluidic flow cells as described in our previous publications.\cite{Wang2019f} Briefly, inlet and outlet ports for electrolyte flow through the cell were achieved by drilling holes in the ITO substrate and inserting Teflon tubing into the holes. Two pieces of approximately 50~µm-thick polytetrafluorethylene (PTFE) spacers were placed between the ITO substrate and a glass cover slip (Thermo Scientific), which formed a microfluidic channel. An electrode chamber containing an Ag/AgI reference electrode and Pt counter electrode was attached to the outlet port. All components were attached/sealed with Loctite epoxy. Electrolyte solution (1M~NaI, Fisher Scientific) was pulled through the cell at a constant rate of 0.5~mL/hr using a Kent Scientific syringe pump. 

\subsection{Photoelectrochemical Measurements}\label{app:echemMeasure}
\vspace{-10pt}
All photoelectrochemical measurements were made using an Ivium Compactstat Potentiostat in a 3-electrode configuration. We assigned anodic photocurrent due to iodide oxidation at the working electrode (the \ce{MoS2}-ITO electrode) to positive current. The open circuit potential of the \ce{MoS2|I^- (1M)} cell was 0.27~V vs. Ag/AgI. For a typical potential-dependent absorbance measurement, the working electrode was held at a fixed potential versus the Ag/AgI reference electrode while \textit{in situ} absorbance spectra were acquired. The potential was stepped from 0.000~V to 0.550~V vs. Ag/AgI in steps of 0.025~V. Likewise, photocurrent spectra were acquired by applying a fixed potential to the working electrode and measuring the resulting current as a function of monochromatic illumination wavelength using a Horiba OBB Tunable PowerArc Illuminator. The light source was scanned from 750~nm to 375~nm in 1~nm steps. The current and monochromator output were synced in time using a data acquisition card (DATAQ Instruments). Monochromatic current-voltage curves were measured from 0.00 V to 0.55 V vs. Ag/AgI while monochromatic light resonant with the A-, B-, or C-excitons (650~nm, 605~nm, and 432~nm, respectively) illuminated the sample. In this case, the light source was chopped at 10~Hz and the current output from the potentiostat was connected to the input channel of a Stanford Research Systems SR830 lock-in amplifier. The lock-in signal was converted from arbitrary units to units of amperes via a proportionality constant, as discussed in our previous work.\cite{Green2003} We calculated the external quantum efficiency (EQE) according to $EQE(\lambda)=qi/I_0(\lambda)$, where $q$ is the electronic charge (in units of~C), $i$ is the photocurrent (in units of~A), and $I_0$ is the monochromatic light power measured at the cover glass (in units of~$s^{–1}$).

\vspace{-16pt}
\subsection{BSE-GW DFT computation}\label{app:bse}
\vspace{-10pt}
The many-body physics of 2D materials in general, and \ce{MoS2} in particular, is computationally challenging. Here we provide details on our calculations. We performed local BSE calculations to obtain the excitonic wavefunctions and energies. Since the primary effect of the converged GW correction is the opening of the band gap, with only minor changes to the dispersion, we used a DFT basis for our BSE corrections at the LDA level with a scissor operator\cite{Baraff1984} to rigidly open the band-gap (i.e. just a constant offset to the conduction bands) to match that in Ref.~\onlinecite{Klots2014}.

\begin{figure}[!t]
\begin{center}
    \resizebox{.33\textwidth}{!}{\includegraphics{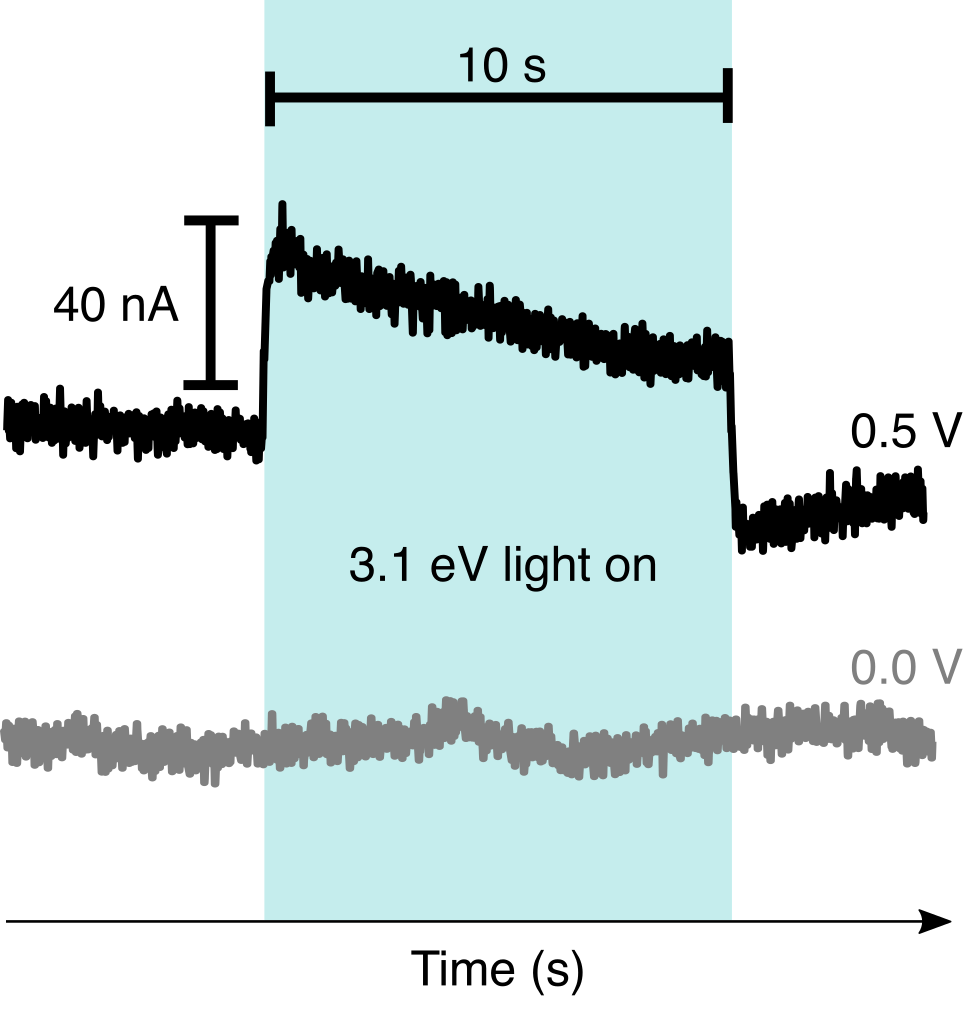}}
\end{center}
\vspace{-14pt}
\caption{\label{fig:si3} Photocurrent onset with pump pulse excitation. Observation of a potential-dependent photocurrent generated by the transient absorption pump laser (3.1 eV).}
\end{figure}

\begin{figure}[!h]
\begin{center}
    \resizebox{.375\textwidth}{!}{\includegraphics{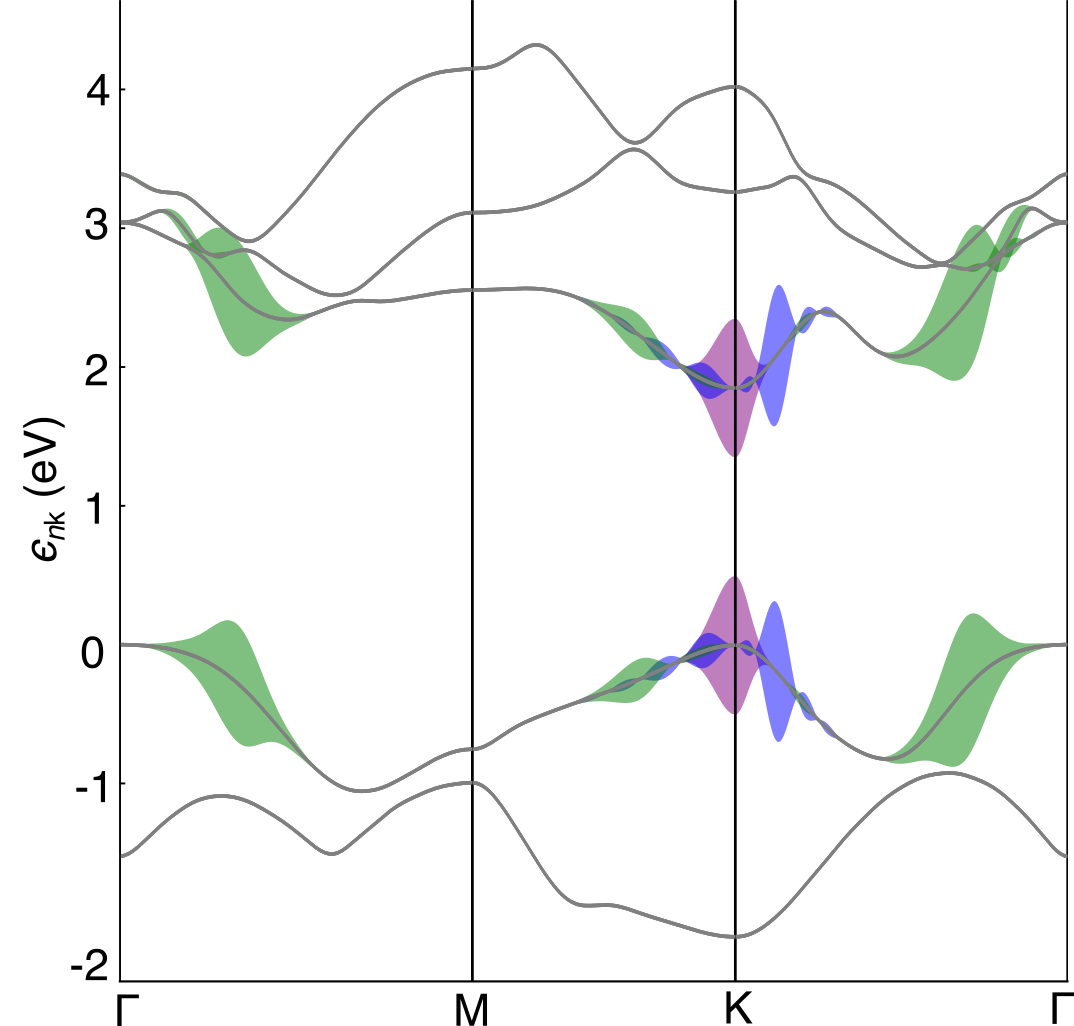}}
\end{center}
\vspace{-14pt}
\caption{\label{fig:si8} Band structure calculated from DFT at the LDA level in a triangular path between the high-symmetry points of the 2D hexagonal Brillouin zone. The colored regions represent the composition of three representative BSE eigenfunctions (after $\sim 1$~eV scissor). The amplitudes are proportional to the weight of that band, at that k point, to the eigenfunction. \textbf{Purple}: lowest energy (1s) A-exciton, \textbf{Green}: lowest energy C-exciton, defined as the first state with amplitude near the $\Gamma$ point, \textbf{Blue}: the state before the first C-exciton, a high-energy A-exciton. The green and blue regions overlap either side of the Kappa point.}
\vspace{-12pt}
\end{figure}

\begin{figure*}[!ht]
\vspace{-12pt}
\begin{center}
    \resizebox{.625\textwidth}{!}{\includegraphics{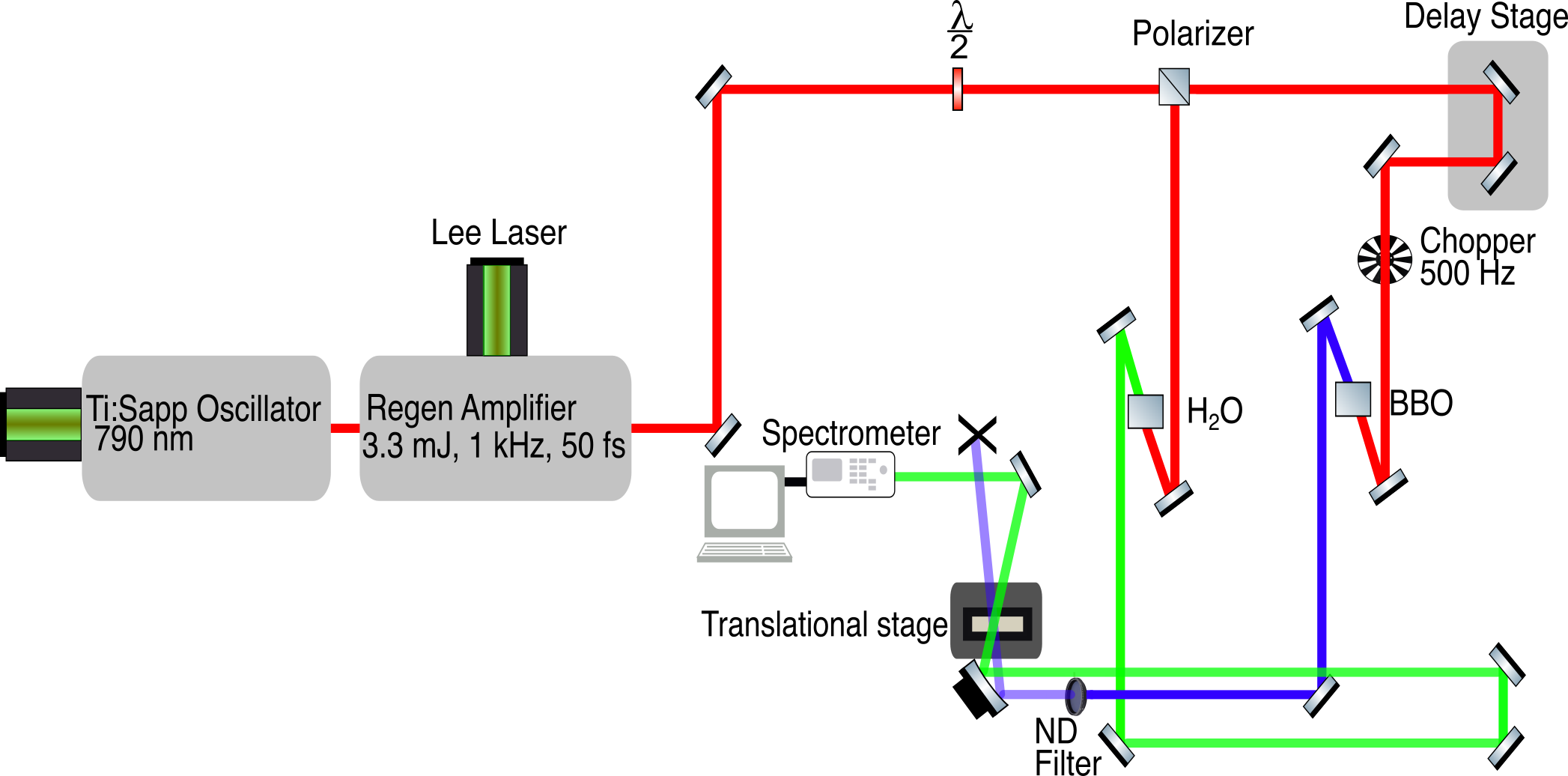}}
\end{center}
\vspace{-14pt}
\caption{\label{fig:si7} Laser table diagram of transient absorption spectrometer. }
\vspace{-8pt}
\end{figure*}

We performed our DFT calculations in Q\textsc{uantum} ESPRESSO\cite{Giannozzi2017} using a relaxed 3-atom cell with: the basal lattice constant at 5.90 Bohr and a sulfur-sulfur distance of 5.79 Bohr. Our calculations show that the position of the conduction band minimum depends strongly on the S--S distance parameter. For example, an unrelaxed value can cause LDA to predict an indirect gap. Other parameters in the calculation included a 43.785 Bohr cell dimension perpendicular to the slab, an energy cutoff of 80~Ry, and a 24×24×1 k-point grid. Trouiller-Martins-type, LDA Ceperley/Alder Perdew/Wang (1992), scalar pseudopotentials were used for the Mo and S, as distributed with the Yambo code.\cite{Sangalli2019} Use of scalar potentials neglects spin-orbit coupling, which results in a faster calculation by two orders of magnitude at the cost of the B-exciton peaks in the structure, and some change to the valence band-edge dispersion.\cite{Molina-Sanchez2013a} The resulting band-structure is shown in Appendix Fig.~\ref{fig:si8}.

We then performed BSE calculations using the Yambo package under the GW approximation,\cite{Sangalli2019} employing screened (static) exchange and a coulomb cutoff equal to 41.785 Bohr and including: two conduction and two valence bands, local fields up to 8~Ry, up to 50 Ry of G-vectors, and 70~bands in the construction of the RPA polarizability. Figure~\ref{fig:2} of the main text shows the optical and eigenspectra arising from this calculation. The isosurface plots of Fig.~\ref{fig:2} were produced using Yambo’s ypp routines, considering 5×5 unit cell repeats and averaging over all possible positions of the exciton’s hole.

\begin{figure}[!t]
\vspace{-1pt}
\begin{center}
    \resizebox{.3\textwidth}{!}{\includegraphics{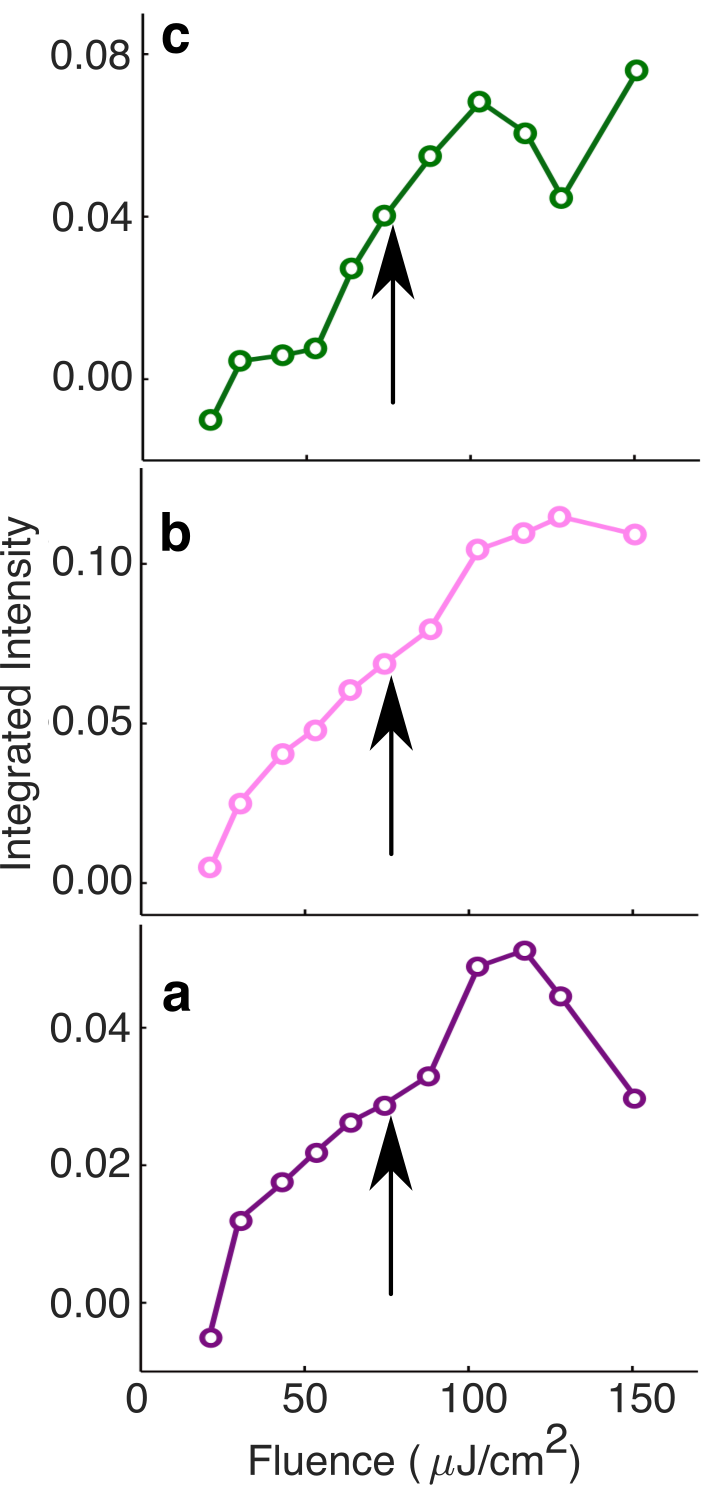}}
\end{center}
\vspace{-16pt}
\caption{\label{fig:si2} Transient absorption pump fluence dependence. Integrated TA bleach intensity (at a probe delay of 1 ps) versus pump fluence for the \textbf{(a)} A-exciton, \textbf{(b)} B-exciton, and \textbf{(c)} C-exciton.  The black arrow highlights the pump fluence of 75 $\upmu$J/cm2 that was chosen for all in situ TA data shown in the main text. The TA bleach intensities are non-linear with pump fluences when the pump fluence exceeds 100 $\upmu$J/cm2, in agreement with literature.\cite{Wang2017e}}
\vspace{-12pt}
\end{figure}

While these parameters seem to converge well and do not present artifacts that would affect the conclusions we draw from the present calculations (e.g., relating the BSE and independent particle absorption peaks), it should be emphasized that they are underconverged. For example, we find an A-exciton binding energy around 1.0~eV, similar to that of Qiu.\cite{Qiu2013} An erratum to that publication\cite{Qiu2015} as well as a recent study\cite{Ataei2021} have both shown that the A-exciton binding energy will continue to decrease up to a k-point grid size of 42×42×1; these are too expensive for our local calculations. This leads to large reduction in the binding energies from their 24×24×1 value, down to around 650\cite{Qiu2015}--680\cite{Ataei2021}~meV. Indeed, this spread of calculated binding energy values is reflected in earlier reviews of the literature.\cite{Berkelbach2018b} However, these quantitative differences should be considered along with the inclusion of other significant factors, such as: the screening of the environment, the phonon coupling, and the conduction band doping (see main text). As such, these absolute values should not be interpreted as being predictive of an experimental setup. The agreement of our calculated A-exciton peak’s position with that of our experiments is due to a cancellation of errors. While we do not expect our qualitative arguments based on the general results of this formalism to be in error based on what is currently known, it is entirely possible that further study into the underlying dynamics of the C-exciton could uncover new physics which qualitatively alters this picture.

\vspace{-14pt}
\subsection{Transient Absorption (TA) Measurements}\label{app:taMeasure}
\vspace{-10pt}
Appendix Fig.~\ref{fig:si7} shows the schematic layout of the transient absorption spectrometer. Transient absorption measurements were performed with a home-built, femtosecond, pump-probe spectrometer. Compressed light from a Ti:sapphire regenerative amplifier (Wyvern 1000, KM Laboratories) produced sub-50 fs pulses centered at 790 nm carrying 3.3 mJ of energy per pulse at a repetition rate of 1kHz. Front reflections off a beam splitter and wedged \ce{CaF2} plate were used to lower the light intensity before splitting the light with a half-wave plate ($\lambda$/2) and polarizer.  A fraction of the light was sent through a delay stage controlled by a Newport motion controller driver (XPS model) to control timing and then directed onto a 1 mm beta barium borate (BBO) crystal to frequency double the light by second harmonic generation to 395 nm. The 395 nm light served as the pump pulse in TA measurements.

\begin{table*}[!htb]
\centering
\begin{tabular}{|l||c|llllll|}
\hline
Exciton & Potential & 0.0 & 0.1 & 0.2 & 0.3 & 0.4 & 0.5 \\ 
& (vs. Ag/AgI) & /V& /V& /V& /V& /V& /V\\\hline\hline
A-exciton & $\tau_1$ /ps & 0.31 ± 0.05 & 0.24 ± 0.01 & 0.310 ± 0.004 & 0.22 ± 0.03 & 0.18 ± 0.01 & 0.210 ± 0.004 \\
 & $\tau_2$	/ps & 3.5 ± 1.0 & 3.7 ± 0.1 & 11.00 ± 0.07 & 4 ± 2 & 10.0 ± 0.7 & 9.5 ± 0.7 \\
 & $\tau_3$	/ps & 22 ± 2 & 45 ± 8 & 130 ± 30 & 82 ± 10 & 260 ± 150 & 220 ± 40 \\
 & $\tau_4$	/ps & 1.1 ± 0.3 & 0.8 ± 0.2 & 0.49 ± 0.04 & 0.64 ± 0.06 & 0.36 ± 0.02 & 0.54 ± 0.05 \\ \hline
B-exciton & $\tau_1$ /ps & 0.24 ± 0.04 & 0.35 ± 0.02 & 0.30 ± 0.03 & 0.22 ± 0.02 & 0.200 ± 0.002 & 0.23 ± 0.01 \\
 & $\tau_2$	/ps & 4 ± 1 & 7.20 ± 0.02 & 8 ± 2 & 10 ± 2 & 9.0 ± 1.0 & 6.5 ± 1.0 \\
 & $\tau_3$	/ps & 61 ± 16 & 65 ± 24 & 45 ± 4 & 70 ± 27 & 250 ± 170 & 67 ± 8 \\
 & $\tau_4$	/ps & 1.7 ± 0.2 & 1.5 ± 0.2 & 0.47 ± 0.10 & 0.37 ± 0.05 & 0.440 ± 0.002 & 0.660 ± 0.005 \\ \hline
C-exciton & $\tau_1$ /ps & 13 ± 5 & 17 ± 2 & 21 ± 2 & 16 ± 7 & 20 ± 4 & 21 ± 5 \\
 & $\tau_2$	/ps & 82 ± 40 & 150 ± 30 & 150 ± 20 & 120 ± 20 & 380 ± 110 & 150 ± 40 \\ \hline
\end{tabular}
\caption{\label{tab:1}Exponential fit (Eq.~\ref{eq:1} and Eq.~\ref{eq:2}) decay parameters as a function of voltage for each exciton peak.}
\vspace{-8pt}
\end{table*}

\begin{figure}[!b]
\vspace{-12pt}
\begin{center}
    \resizebox{.33\textwidth}{!}{\includegraphics{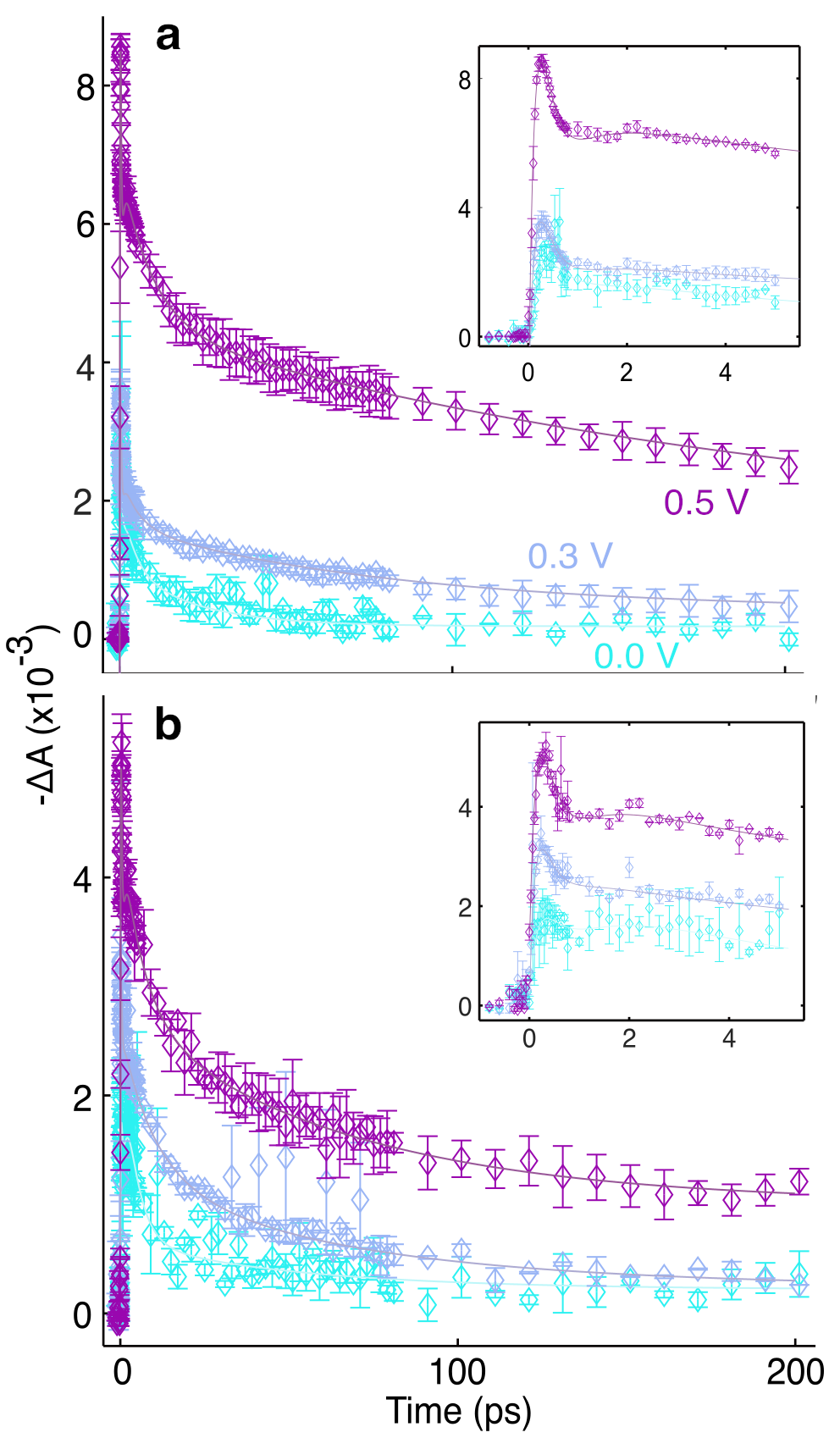}}
\end{center}
\vspace{-14pt}
\caption{\label{fig:si4} Transient absorption A- and B-exciton temporal fits. Tri-exponential fitting of the TA bleach decay curves for the \textbf{(a)} A- and \textbf{(b)} B-excitons at applied potentials of 0.5 V (dark purple), 0.3 V (lavender), and 0.0 V (teal).  The inset shows the first 5 ps of the decay traces. See Appendix Table~\ref{tab:1} for extracted fit time constants.}
\vspace{-12pt}
\end{figure}

A neutral density (ND) filter was used to adjust the pump fluence. The probe pulse was created by taking the remaining portion of the split light and focusing it down in a 2~mm quartz cuvette filled with 18~M$\Omega$ water, creating a white-light continuum spanning 400~nm to 800~nm. The pump and probe pulses were focused to a spot size of approximately 30~$\upmu$m full-width half-max and overlapped at the sample in a non-collinear beam geometry. The probe light transmitted through the sample was collected, collimated, and focused into a spectrometer (iHR550, Horiba) with a 200~mm slit opening, equipped with a 100~line/mm grating (450 blaze), and detected by a single line 2048~element array detector (OctoPlus, Teledyne e2V). The spectral and temporal resolution of the setup was 0.6~nm and $\sim 50$~fs, respectively. The pump pulse was modulated by an optical chopper (Thorlabs) triggered by the pump laser and set to operate at half the repetition rate of the laser (500~Hz).  The spectrally resolved “pumped” and “unpumped” signals are acquired and used to calculate the $\Delta A$ spectrum.  A pump fluence of 75~µJ/cm$^2$ is used throughout this investigation.

\subsection{TA Temporal Decay Analysis}\label{app:decay}
\vspace{-10pt}
Following literature,\cite{Trovatello2020, Shi2013} the A-exciton and B-exciton bleach intensity versus time data were fit using a tri-exponential function convolved with a rising exponential of the form
\begin{equation}\label{eq:1}
    \Delta A (\lambda, t) = A_0 + \left( \sum_{i=1}^3 A_i (\lambda) \exp{\frac{-t}{\tau_i}} \right)\left(  A_4-\exp{\frac{-t}{\tau_4}} \right).
\end{equation}
\vspace{4pt}

The fast, medium, and slow components of the triexponential decay have been assigned to defect and carrier-carrier scattering, carrier-phonon scattering, and electron-hole recombination processes.\cite{Shi2013, DalConte2020} The rising exponential function captures the exciton formation process (Appendix Fig.~\ref{fig:si4}). Fit constants for the A- and B-excitons are in Appendix Table~\ref{tab:1}.\\

\begin{figure}[!b]
\begin{center}
    \resizebox{.33\textwidth}{!}{\includegraphics{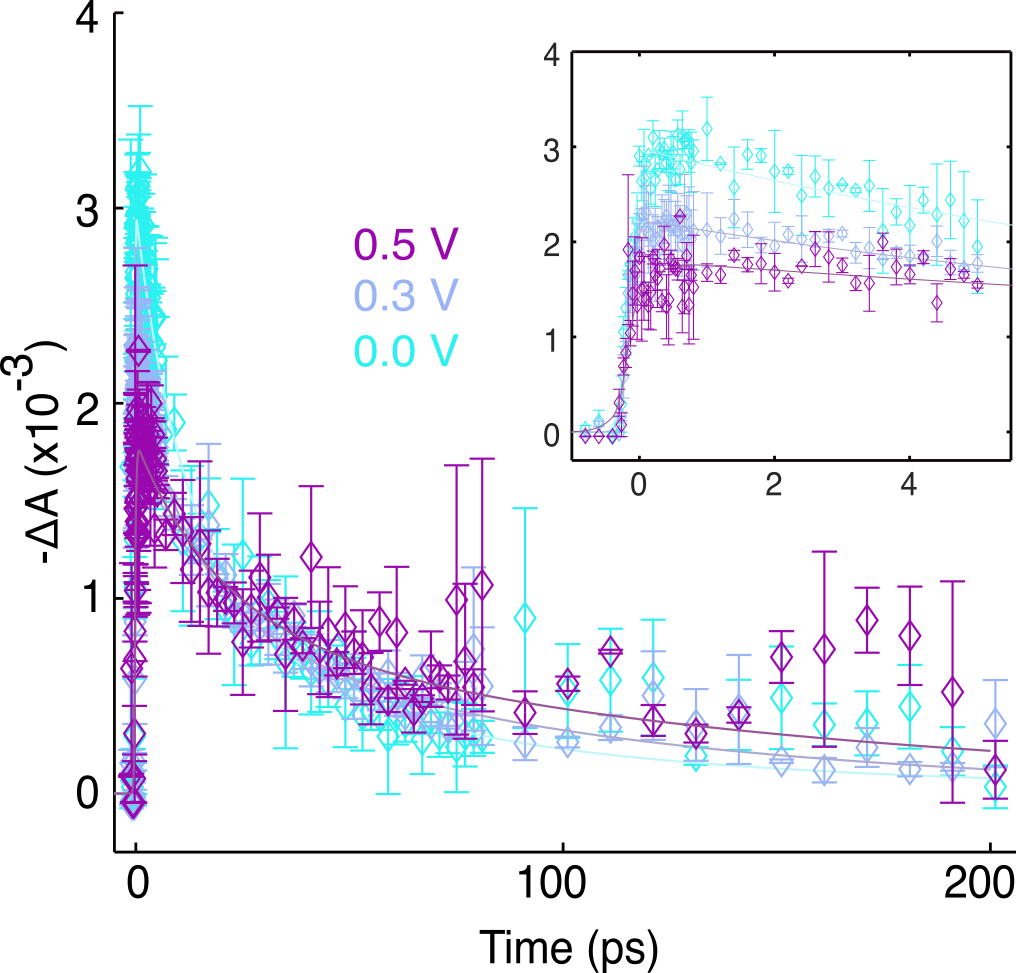}}
\end{center}
\vspace{-14pt}
\caption{\label{fig:si5} Transient absorption decay curves for the C-exciton. Bi-exponential fitting of the TA bleach decay curves at 0.5 V (dark purple), 0.3 V (lavender), and 0.0 V (teal).  The inset shows the first 5~ps of the decay traces. See Appendix Table~\ref{tab:1} for extracted fit time constants.}
\vspace{-3pt}
\end{figure}

The C-exciton TA data (Appendix Fig.~\ref{fig:si5}) could be fit with a biexponential function convolved with the instrument response function of the TA spectrometer,
\begin{align}\label{eq:2}
    \Delta A &(\lambda, t) = A_0 \,\, + \nonumber\\
    &\sum_{i=1}^2 \frac{A_i(\lambda)}{2} \exp{\left( \frac{-1}{\tau_i} \right)\left( t - \frac{\sigma^2}{2\tau_i} \right)} \left( 1- \rmm{erf}\left[ \frac{t-\sigma^2}{\sigma\tau_i \sqrt(2)} \right] \right).
\end{align}
The fast and slow components of the C-exciton decay dynamics have been assigned to carrier-phonon scattering and intervalley relaxation processes. For all fits, a static offset ($A_0$) is used in the functional form to capture long-lived signal that extends beyond the temporal window of our experiment, in agreement with literature.\cite{Ruckebusch2012} Fit constants for the C-exciton are in Table~\ref{tab:1}.

\vfill\pagebreak
\subsection*{References}
\vspace{-16pt}
\bibliography{Postdoc-TMDC}

\begin{thebibliography}{57}%
\makeatletter
\providecommand \@ifxundefined [1]{%
 \@ifx{#1\undefined}
}%
\providecommand \@ifnum [1]{%
 \ifnum #1\expandafter \@firstoftwo
 \else \expandafter \@secondoftwo
 \fi
}%
\providecommand \@ifx [1]{%
 \ifx #1\expandafter \@firstoftwo
 \else \expandafter \@secondoftwo
 \fi
}%
\providecommand \natexlab [1]{#1}%
\providecommand \enquote  [1]{``#1''}%
\providecommand \bibnamefont  [1]{#1}%
\providecommand \bibfnamefont [1]{#1}%
\providecommand \citenamefont [1]{#1}%
\providecommand \href@noop [0]{\@secondoftwo}%
\providecommand \href [0]{\begingroup \@sanitize@url \@href}%
\providecommand \@href[1]{\@@startlink{#1}\@@href}%
\providecommand \@@href[1]{\endgroup#1\@@endlink}%
\providecommand \@sanitize@url [0]{\catcode `\\12\catcode `\$12\catcode
  `\&12\catcode `\#12\catcode `\^12\catcode `\_12\catcode `\%12\relax}%
\providecommand \@@startlink[1]{}%
\providecommand \@@endlink[0]{}%
\providecommand \url  [0]{\begingroup\@sanitize@url \@url }%
\providecommand \@url [1]{\endgroup\@href {#1}{\urlprefix }}%
\providecommand \urlprefix  [0]{URL }%
\providecommand \Eprint [0]{\href }%
\providecommand \doibase [0]{http://dx.doi.org/}%
\providecommand \selectlanguage [0]{\@gobble}%
\providecommand \bibinfo  [0]{\@secondoftwo}%
\providecommand \bibfield  [0]{\@secondoftwo}%
\providecommand \translation [1]{[#1]}%
\providecommand \BibitemOpen [0]{}%
\providecommand \bibitemStop [0]{}%
\providecommand \bibitemNoStop [0]{.\EOS\space}%
\providecommand \EOS [0]{\spacefactor3000\relax}%
\providecommand \BibitemShut  [1]{\csname bibitem#1\endcsname}%
\let\auto@bib@innerbib\@empty
\bibitem [{\citenamefont {Nozik}(2003)}]{Nozik2003}%
  \BibitemOpen
  \bibfield  {author} {\bibinfo {author} {\bibfnamefont {A.~J.}\ \bibnamefont
  {Nozik}},\ }\bibfield  {title} {\enquote {\bibinfo {title} {{Spectroscopy and
  Hot Electron Relaxation Dynamics in Semiconductor Quantum Wells and Quantum
  Dots}},}\ }\href {\doibase 10.1146/ANNUREV.PHYSCHEM.52.1.193} {\bibfield
  {journal} {\bibinfo  {journal} {Annual Review of Physical Chemistry}\
  }\textbf {\bibinfo {volume} {52}},\ \bibinfo {pages} {193--231} (\bibinfo
  {year} {2003})}\BibitemShut {NoStop}%
\bibitem [{\citenamefont {Shockley}\ and\ \citenamefont
  {Queisser}(2004)}]{Shockley2004}%
  \BibitemOpen
  \bibfield  {author} {\bibinfo {author} {\bibfnamefont {W.}~\bibnamefont
  {Shockley}}\ and\ \bibinfo {author} {\bibfnamefont {H.~J.}\ \bibnamefont
  {Queisser}},\ }\bibfield  {title} {\enquote {\bibinfo {title} {{Detailed
  Balance Limit of Efficiency of p‐n Junction Solar Cells}},}\ }\href
  {\doibase 10.1063/1.1736034} {\bibfield  {journal} {\bibinfo  {journal}
  {Journal of Applied Physics}\ }\textbf {\bibinfo {volume} {32}},\ \bibinfo
  {pages} {510} (\bibinfo {year} {2004})}\BibitemShut {NoStop}%
\bibitem [{\citenamefont {Nozik}(2018)}]{Nozik2018}%
  \BibitemOpen
  \bibfield  {author} {\bibinfo {author} {\bibfnamefont {A.~J.}\ \bibnamefont
  {Nozik}},\ }\bibfield  {title} {\enquote {\bibinfo {title} {{Utilizing hot
  electrons}},}\ }\href {\doibase 10.1038/s41560-018-0112-5} {\bibfield
  {journal} {\bibinfo  {journal} {Nature Energy}\ }\textbf {\bibinfo {volume}
  {3}},\ \bibinfo {pages} {170--171} (\bibinfo {year} {2018})}\BibitemShut
  {NoStop}%
\bibitem [{\citenamefont {Cooper}\ \emph {et~al.}(1998)\citenamefont {Cooper},
  \citenamefont {Turner}, \citenamefont {Parkinson},\ and\ \citenamefont
  {Nozik}}]{Cooper1998}%
  \BibitemOpen
  \bibfield  {author} {\bibinfo {author} {\bibfnamefont {G.}~\bibnamefont
  {Cooper}}, \bibinfo {author} {\bibfnamefont {J.~A.}\ \bibnamefont {Turner}},
  \bibinfo {author} {\bibfnamefont {B.~A.}\ \bibnamefont {Parkinson}}, \ and\
  \bibinfo {author} {\bibfnamefont {A.~J.}\ \bibnamefont {Nozik}},\ }\bibfield
  {title} {\enquote {\bibinfo {title} {{Hot carrier injection of photogenerated
  electrons at indium phosphide–electrolyte interfaces}},}\ }\href {\doibase
  10.1063/1.331928} {\bibfield  {journal} {\bibinfo  {journal} {Journal of
  Applied Physics}\ }\textbf {\bibinfo {volume} {54}},\ \bibinfo {pages} {6463}
  (\bibinfo {year} {1998})}\BibitemShut {NoStop}%
\bibitem [{\citenamefont {Nguyen}\ \emph {et~al.}(2018)\citenamefont {Nguyen},
  \citenamefont {Lombez}, \citenamefont {Gibelli}, \citenamefont
  {Boyer-Richard}, \citenamefont {{Le Corre}}, \citenamefont {Durand},\ and\
  \citenamefont {Guillemoles}}]{Nguyen2018}%
  \BibitemOpen
  \bibfield  {author} {\bibinfo {author} {\bibfnamefont {D.~T.}\ \bibnamefont
  {Nguyen}}, \bibinfo {author} {\bibfnamefont {L.}~\bibnamefont {Lombez}},
  \bibinfo {author} {\bibfnamefont {F.}~\bibnamefont {Gibelli}}, \bibinfo
  {author} {\bibfnamefont {S.}~\bibnamefont {Boyer-Richard}}, \bibinfo {author}
  {\bibfnamefont {A.}~\bibnamefont {{Le Corre}}}, \bibinfo {author}
  {\bibfnamefont {O.}~\bibnamefont {Durand}}, \ and\ \bibinfo {author}
  {\bibfnamefont {J.~F.}\ \bibnamefont {Guillemoles}},\ }\bibfield  {title}
  {\enquote {\bibinfo {title} {{Quantitative experimental assessment of hot
  carrier-enhanced solar cells at room temperature}},}\ }\href {\doibase
  10.1038/s41560-018-0106-3} {\bibfield  {journal} {\bibinfo  {journal} {Nature
  Energy}\ }\textbf {\bibinfo {volume} {3}},\ \bibinfo {pages} {236--242}
  (\bibinfo {year} {2018})}\BibitemShut {NoStop}%
\bibitem [{\citenamefont {Nozik}(2021)}]{Nozik2021a}%
  \BibitemOpen
  \bibfield  {author} {\bibinfo {author} {\bibfnamefont {A.~J.}\ \bibnamefont
  {Nozik}},\ }\bibfield  {title} {\enquote {\bibinfo {title} {{Quantization
  effects in semiconductor nanostructures and singlet fission in molecular
  chromophores for photovoltaics and solar fuels}},}\ }\href {\doibase
  10.1063/5.0028982} {\bibfield  {journal} {\bibinfo  {journal} {Chemical
  Physics Reviews}\ }\textbf {\bibinfo {volume} {2}},\ \bibinfo {pages}
  {021305} (\bibinfo {year} {2021})}\BibitemShut {NoStop}%
\bibitem [{\citenamefont {Gabor}\ \emph {et~al.}(2011)\citenamefont {Gabor},
  \citenamefont {Song}, \citenamefont {Ma}, \citenamefont {Nair}, \citenamefont
  {Taychatanapat}, \citenamefont {Watanabe}, \citenamefont {Taniguchi},
  \citenamefont {Levitov},\ and\ \citenamefont {Jarillo-Herrero}}]{Gabor2011}%
  \BibitemOpen
  \bibfield  {author} {\bibinfo {author} {\bibfnamefont {N.~M.}\ \bibnamefont
  {Gabor}}, \bibinfo {author} {\bibfnamefont {J.~C.}\ \bibnamefont {Song}},
  \bibinfo {author} {\bibfnamefont {Q.}~\bibnamefont {Ma}}, \bibinfo {author}
  {\bibfnamefont {N.~L.}\ \bibnamefont {Nair}}, \bibinfo {author}
  {\bibfnamefont {T.}~\bibnamefont {Taychatanapat}}, \bibinfo {author}
  {\bibfnamefont {K.}~\bibnamefont {Watanabe}}, \bibinfo {author}
  {\bibfnamefont {T.}~\bibnamefont {Taniguchi}}, \bibinfo {author}
  {\bibfnamefont {L.~S.}\ \bibnamefont {Levitov}}, \ and\ \bibinfo {author}
  {\bibfnamefont {P.}~\bibnamefont {Jarillo-Herrero}},\ }\bibfield  {title}
  {\enquote {\bibinfo {title} {{Hot carrier-assisted intrinsic photoresponse in
  graphene}},}\ }\href {\doibase
  10.1126/SCIENCE.1211384/SUPPL_FILE/GABOR.SOM.PDF} {\bibfield  {journal}
  {\bibinfo  {journal} {Science}\ }\textbf {\bibinfo {volume} {334}},\ \bibinfo
  {pages} {648--652} (\bibinfo {year} {2011})}\BibitemShut {NoStop}%
\bibitem [{\citenamefont {Sun}\ \emph {et~al.}(2012)\citenamefont {Sun},
  \citenamefont {Aivazian}, \citenamefont {Jones}, \citenamefont {Ross},
  \citenamefont {Yao}, \citenamefont {Cobden},\ and\ \citenamefont
  {Xu}}]{Sun2012a}%
  \BibitemOpen
  \bibfield  {author} {\bibinfo {author} {\bibfnamefont {D.}~\bibnamefont
  {Sun}}, \bibinfo {author} {\bibfnamefont {G.}~\bibnamefont {Aivazian}},
  \bibinfo {author} {\bibfnamefont {A.~M.}\ \bibnamefont {Jones}}, \bibinfo
  {author} {\bibfnamefont {J.~S.}\ \bibnamefont {Ross}}, \bibinfo {author}
  {\bibfnamefont {W.}~\bibnamefont {Yao}}, \bibinfo {author} {\bibfnamefont
  {D.}~\bibnamefont {Cobden}}, \ and\ \bibinfo {author} {\bibfnamefont
  {X.}~\bibnamefont {Xu}},\ }\bibfield  {title} {\enquote {\bibinfo {title}
  {{Ultrafast hot-carrier-dominated photocurrent in graphene}},}\ }\href
  {\doibase 10.1038/nnano.2011.243} {\bibfield  {journal} {\bibinfo  {journal}
  {Nature Nanotechnology}\ }\textbf {\bibinfo {volume} {7}},\ \bibinfo {pages}
  {114--118} (\bibinfo {year} {2012})}\BibitemShut {NoStop}%
\bibitem [{\citenamefont {Xu}\ \emph {et~al.}(2010)\citenamefont {Xu},
  \citenamefont {Gabor}, \citenamefont {Alden}, \citenamefont {{Van Der
  Zande}},\ and\ \citenamefont {McEuen}}]{Xu2010b}%
  \BibitemOpen
  \bibfield  {author} {\bibinfo {author} {\bibfnamefont {X.}~\bibnamefont
  {Xu}}, \bibinfo {author} {\bibfnamefont {N.~M.}\ \bibnamefont {Gabor}},
  \bibinfo {author} {\bibfnamefont {J.~S.}\ \bibnamefont {Alden}}, \bibinfo
  {author} {\bibfnamefont {A.~M.}\ \bibnamefont {{Van Der Zande}}}, \ and\
  \bibinfo {author} {\bibfnamefont {P.~L.}\ \bibnamefont {McEuen}},\ }\bibfield
   {title} {\enquote {\bibinfo {title} {{Photo-thermoelectric effect at a
  graphene interface junction}},}\ }\href {\doibase
  10.1021/NL903451Y/ASSET/IMAGES/LARGE/NL-2009-03451Y_0005.JPEG} {\bibfield
  {journal} {\bibinfo  {journal} {Nano Letters}\ }\textbf {\bibinfo {volume}
  {10}},\ \bibinfo {pages} {562--566} (\bibinfo {year} {2010})}\BibitemShut
  {NoStop}%
\bibitem [{\citenamefont {Zhang}, \citenamefont {Yam},\ and\ \citenamefont
  {Schatz}(2016)}]{Zhang2016d}%
  \BibitemOpen
  \bibfield  {author} {\bibinfo {author} {\bibfnamefont {Y.}~\bibnamefont
  {Zhang}}, \bibinfo {author} {\bibfnamefont {C.}~\bibnamefont {Yam}}, \ and\
  \bibinfo {author} {\bibfnamefont {G.~C.}\ \bibnamefont {Schatz}},\ }\bibfield
   {title} {\enquote {\bibinfo {title} {{Fundamental Limitations to Plasmonic
  Hot-Carrier Solar Cells}},}\ }\href {\doibase
  10.1021/ACS.JPCLETT.6B00879/ASSET/IMAGES/LARGE/JZ-2016-00879H_0006.JPEG}
  {\bibfield  {journal} {\bibinfo  {journal} {Journal of Physical Chemistry
  Letters}\ }\textbf {\bibinfo {volume} {7}},\ \bibinfo {pages} {1852--1858}
  (\bibinfo {year} {2016})}\BibitemShut {NoStop}%
\bibitem [{\citenamefont {Li}\ \emph {et~al.}(2017)\citenamefont {Li},
  \citenamefont {Bhaumik}, \citenamefont {Goh}, \citenamefont {Kumar},
  \citenamefont {Yantara}, \citenamefont {Gr{\"{a}}tzel}, \citenamefont
  {Mhaisalkar}, \citenamefont {Mathews},\ and\ \citenamefont {Sum}}]{Li2017a}%
  \BibitemOpen
  \bibfield  {author} {\bibinfo {author} {\bibfnamefont {M.}~\bibnamefont
  {Li}}, \bibinfo {author} {\bibfnamefont {S.}~\bibnamefont {Bhaumik}},
  \bibinfo {author} {\bibfnamefont {T.~W.}\ \bibnamefont {Goh}}, \bibinfo
  {author} {\bibfnamefont {M.~S.}\ \bibnamefont {Kumar}}, \bibinfo {author}
  {\bibfnamefont {N.}~\bibnamefont {Yantara}}, \bibinfo {author} {\bibfnamefont
  {M.}~\bibnamefont {Gr{\"{a}}tzel}}, \bibinfo {author} {\bibfnamefont
  {S.}~\bibnamefont {Mhaisalkar}}, \bibinfo {author} {\bibfnamefont
  {N.}~\bibnamefont {Mathews}}, \ and\ \bibinfo {author} {\bibfnamefont
  {T.~C.}\ \bibnamefont {Sum}},\ }\bibfield  {title} {\enquote {\bibinfo
  {title} {{Slow cooling and highly efficient extraction of hot carriers in
  colloidal perovskite nanocrystals}},}\ }\href {\doibase 10.1038/ncomms14350}
  {\bibfield  {journal} {\bibinfo  {journal} {Nature Communications}\ }\textbf
  {\bibinfo {volume} {8}},\ \bibinfo {pages} {1--10} (\bibinfo {year}
  {2017})}\BibitemShut {NoStop}%
\bibitem [{\citenamefont {Tisdale}\ \emph {et~al.}(2010)\citenamefont
  {Tisdale}, \citenamefont {Williams}, \citenamefont {Timp}, \citenamefont
  {Norris}, \citenamefont {Aydil},\ and\ \citenamefont {Zhu}}]{Tisdale2010}%
  \BibitemOpen
  \bibfield  {author} {\bibinfo {author} {\bibfnamefont {W.~A.}\ \bibnamefont
  {Tisdale}}, \bibinfo {author} {\bibfnamefont {K.~J.}\ \bibnamefont
  {Williams}}, \bibinfo {author} {\bibfnamefont {B.~A.}\ \bibnamefont {Timp}},
  \bibinfo {author} {\bibfnamefont {D.~J.}\ \bibnamefont {Norris}}, \bibinfo
  {author} {\bibfnamefont {E.~S.}\ \bibnamefont {Aydil}}, \ and\ \bibinfo
  {author} {\bibfnamefont {X.~Y.}\ \bibnamefont {Zhu}},\ }\bibfield  {title}
  {\enquote {\bibinfo {title} {{Hot-electron transfer from semiconductor
  nanocrystals}},}\ }\href {\doibase
  10.1126/SCIENCE.1185509/SUPPL_FILE/TISDALE.SOM.PDF} {\bibfield  {journal}
  {\bibinfo  {journal} {Science}\ }\textbf {\bibinfo {volume} {328}},\ \bibinfo
  {pages} {1543--1547} (\bibinfo {year} {2010})}\BibitemShut {NoStop}%
\bibitem [{\citenamefont {Verkamp}\ \emph {et~al.}(2021)\citenamefont
  {Verkamp}, \citenamefont {Leveillee}, \citenamefont {Sharma}, \citenamefont
  {Lin}, \citenamefont {Schleife},\ and\ \citenamefont
  {Vura-Weis}}]{Verkamp2021}%
  \BibitemOpen
  \bibfield  {author} {\bibinfo {author} {\bibfnamefont {M.}~\bibnamefont
  {Verkamp}}, \bibinfo {author} {\bibfnamefont {J.}~\bibnamefont {Leveillee}},
  \bibinfo {author} {\bibfnamefont {A.}~\bibnamefont {Sharma}}, \bibinfo
  {author} {\bibfnamefont {M.~F.}\ \bibnamefont {Lin}}, \bibinfo {author}
  {\bibfnamefont {A.}~\bibnamefont {Schleife}}, \ and\ \bibinfo {author}
  {\bibfnamefont {J.}~\bibnamefont {Vura-Weis}},\ }\bibfield  {title} {\enquote
  {\bibinfo {title} {{Carrier-Specific Hot Phonon Bottleneck in CH3NH3PbI3
  Revealed by Femtosecond XUV Absorption}},}\ }\href {\doibase
  10.1021/JACS.1C07817/ASSET/IMAGES/MEDIUM/JA1C07817_M001.GIF} {\bibfield
  {journal} {\bibinfo  {journal} {Journal of the American Chemical Society}\
  }\textbf {\bibinfo {volume} {143}},\ \bibinfo {pages} {20176--20182}
  (\bibinfo {year} {2021})}\BibitemShut {NoStop}%
\bibitem [{\citenamefont {Wadia}, \citenamefont {Alivisatos},\ and\
  \citenamefont {Kammen}(2009)}]{Wadia2009}%
  \BibitemOpen
  \bibfield  {author} {\bibinfo {author} {\bibfnamefont {C.}~\bibnamefont
  {Wadia}}, \bibinfo {author} {\bibfnamefont {A.~P.}\ \bibnamefont
  {Alivisatos}}, \ and\ \bibinfo {author} {\bibfnamefont {D.~M.}\ \bibnamefont
  {Kammen}},\ }\bibfield  {title} {\enquote {\bibinfo {title} {{Materials
  availability expands the opportunity for large-scale photovoltaics
  deployment}},}\ }\href {\doibase
  10.1021/ES8019534/SUPPL_FILE/ES8019534_SI_001.PDF} {\bibfield  {journal}
  {\bibinfo  {journal} {Environmental Science and Technology}\ }\textbf
  {\bibinfo {volume} {43}},\ \bibinfo {pages} {2072--2077} (\bibinfo {year}
  {2009})}\BibitemShut {NoStop}%
\bibitem [{\citenamefont {Kline}\ \emph {et~al.}(1981)\citenamefont {Kline},
  \citenamefont {Kam}, \citenamefont {Canfield},\ and\ \citenamefont
  {Parkinson}}]{Kline1981}%
  \BibitemOpen
  \bibfield  {author} {\bibinfo {author} {\bibfnamefont {G.}~\bibnamefont
  {Kline}}, \bibinfo {author} {\bibfnamefont {K.}~\bibnamefont {Kam}}, \bibinfo
  {author} {\bibfnamefont {D.}~\bibnamefont {Canfield}}, \ and\ \bibinfo
  {author} {\bibfnamefont {B.~A.}\ \bibnamefont {Parkinson}},\ }\bibfield
  {title} {\enquote {\bibinfo {title} {{Efficient and stable
  photoelectrochemical cells constructed with WSe2 and MoSe2 photoanodes}},}\
  }\href {\doibase 10.1016/0165-1633(81)90068-X} {\bibfield  {journal}
  {\bibinfo  {journal} {Solar Energy Materials}\ }\textbf {\bibinfo {volume}
  {4}},\ \bibinfo {pages} {301--308} (\bibinfo {year} {1981})}\BibitemShut
  {NoStop}%
\bibitem [{\citenamefont {Wang}\ \emph {et~al.}(2017)\citenamefont {Wang},
  \citenamefont {Wang}, \citenamefont {Wang}, \citenamefont {Grinblat},
  \citenamefont {Huang}, \citenamefont {Wang}, \citenamefont {Ye},
  \citenamefont {Li}, \citenamefont {Bao}, \citenamefont {Wee}, \citenamefont
  {Maier}, \citenamefont {Chen}, \citenamefont {Zhong}, \citenamefont {Qiu},\
  and\ \citenamefont {Sun}}]{Wang2017e}%
  \BibitemOpen
  \bibfield  {author} {\bibinfo {author} {\bibfnamefont {L.}~\bibnamefont
  {Wang}}, \bibinfo {author} {\bibfnamefont {Z.}~\bibnamefont {Wang}}, \bibinfo
  {author} {\bibfnamefont {H.~Y.}\ \bibnamefont {Wang}}, \bibinfo {author}
  {\bibfnamefont {G.}~\bibnamefont {Grinblat}}, \bibinfo {author}
  {\bibfnamefont {Y.~L.}\ \bibnamefont {Huang}}, \bibinfo {author}
  {\bibfnamefont {D.}~\bibnamefont {Wang}}, \bibinfo {author} {\bibfnamefont
  {X.~H.}\ \bibnamefont {Ye}}, \bibinfo {author} {\bibfnamefont {X.~B.}\
  \bibnamefont {Li}}, \bibinfo {author} {\bibfnamefont {Q.}~\bibnamefont
  {Bao}}, \bibinfo {author} {\bibfnamefont {A.~S.}\ \bibnamefont {Wee}},
  \bibinfo {author} {\bibfnamefont {S.~A.}\ \bibnamefont {Maier}}, \bibinfo
  {author} {\bibfnamefont {Q.~D.}\ \bibnamefont {Chen}}, \bibinfo {author}
  {\bibfnamefont {M.~L.}\ \bibnamefont {Zhong}}, \bibinfo {author}
  {\bibfnamefont {C.~W.}\ \bibnamefont {Qiu}}, \ and\ \bibinfo {author}
  {\bibfnamefont {H.~B.}\ \bibnamefont {Sun}},\ }\bibfield  {title} {\enquote
  {\bibinfo {title} {{Slow cooling and efficient extraction of C-exciton hot
  carriers in MoS2 monolayer}},}\ }\href {\doibase 10.1038/ncomms13906}
  {\bibfield  {journal} {\bibinfo  {journal} {Nature Communications}\ }\textbf
  {\bibinfo {volume} {8}},\ \bibinfo {pages} {1--8} (\bibinfo {year}
  {2017})}\BibitemShut {NoStop}%
\bibitem [{\citenamefont {Rose}\ \emph {et~al.}(2022)\citenamefont {Rose},
  \citenamefont {Aubry}, \citenamefont {Zhang}, \citenamefont {Vigil-Fowler},\
  and\ \citenamefont {{Van De Lagemaat}}}]{Rose2022}%
  \BibitemOpen
  \bibfield  {author} {\bibinfo {author} {\bibfnamefont {A.~H.}\ \bibnamefont
  {Rose}}, \bibinfo {author} {\bibfnamefont {T.~J.}\ \bibnamefont {Aubry}},
  \bibinfo {author} {\bibfnamefont {H.}~\bibnamefont {Zhang}}, \bibinfo
  {author} {\bibfnamefont {D.}~\bibnamefont {Vigil-Fowler}}, \ and\ \bibinfo
  {author} {\bibfnamefont {J.}~\bibnamefont {{Van De Lagemaat}}},\ }\bibfield
  {title} {\enquote {\bibinfo {title} {{Ultrastrong Coupling Leads to Slowed
  Cooling of Hot Excitons in Few-Layer Transition-Metal Dichalcogenides}},}\
  }\href {\doibase
  10.1021/ACS.JPCC.2C01262/ASSET/IMAGES/LARGE/JP2C01262_0005.JPEG} {\bibfield
  {journal} {\bibinfo  {journal} {Journal of Physical Chemistry C}\ }\textbf
  {\bibinfo {volume} {126}},\ \bibinfo {pages} {8710--8719} (\bibinfo {year}
  {2022})}\BibitemShut {NoStop}%
\bibitem [{\citenamefont {Vogt}\ \emph {et~al.}(2020)\citenamefont {Vogt},
  \citenamefont {Shi}, \citenamefont {Wang},\ and\ \citenamefont
  {Graham}}]{Vogt2020}%
  \BibitemOpen
  \bibfield  {author} {\bibinfo {author} {\bibfnamefont {K.~T.}\ \bibnamefont
  {Vogt}}, \bibinfo {author} {\bibfnamefont {S.~F.}\ \bibnamefont {Shi}},
  \bibinfo {author} {\bibfnamefont {F.}~\bibnamefont {Wang}}, \ and\ \bibinfo
  {author} {\bibfnamefont {M.~W.}\ \bibnamefont {Graham}},\ }\bibfield  {title}
  {\enquote {\bibinfo {title} {{Ultrafast Photocurrent and Absorption
  Microscopy of Few-Layer Transition Metal Dichalcogenide Devices That Isolate
  Rate-Limiting Dynamics Driving Fast and Efficient Photoresponse}},}\ }\href
  {\doibase 10.1021/ACS.JPCC.0C02646/ASSET/IMAGES/LARGE/JP0C02646_0005.JPEG}
  {\bibfield  {journal} {\bibinfo  {journal} {Journal of Physical Chemistry C}\
  }\textbf {\bibinfo {volume} {124}},\ \bibinfo {pages} {15195--15204}
  (\bibinfo {year} {2020})}\BibitemShut {NoStop}%
\bibitem [{\citenamefont {{Grubi{\v{s}}i{\'{c}} {\v{C}}abo}}\ \emph
  {et~al.}(2015)\citenamefont {{Grubi{\v{s}}i{\'{c}} {\v{C}}abo}},
  \citenamefont {Miwa}, \citenamefont {Gr{\o}nborg}, \citenamefont {Riley},
  \citenamefont {Johannsen}, \citenamefont {Cacho}, \citenamefont {Alexander},
  \citenamefont {Chapman}, \citenamefont {Springate}, \citenamefont {Grioni},
  \citenamefont {Lauritsen}, \citenamefont {King}, \citenamefont {Hofmann},\
  and\ \citenamefont {Ulstrup}}]{GrubisicCabo2015}%
  \BibitemOpen
  \bibfield  {author} {\bibinfo {author} {\bibfnamefont {A.}~\bibnamefont
  {{Grubi{\v{s}}i{\'{c}} {\v{C}}abo}}}, \bibinfo {author} {\bibfnamefont
  {J.~A.}\ \bibnamefont {Miwa}}, \bibinfo {author} {\bibfnamefont {S.~S.}\
  \bibnamefont {Gr{\o}nborg}}, \bibinfo {author} {\bibfnamefont {J.~M.}\
  \bibnamefont {Riley}}, \bibinfo {author} {\bibfnamefont {J.~C.}\ \bibnamefont
  {Johannsen}}, \bibinfo {author} {\bibfnamefont {C.}~\bibnamefont {Cacho}},
  \bibinfo {author} {\bibfnamefont {O.}~\bibnamefont {Alexander}}, \bibinfo
  {author} {\bibfnamefont {R.~T.}\ \bibnamefont {Chapman}}, \bibinfo {author}
  {\bibfnamefont {E.}~\bibnamefont {Springate}}, \bibinfo {author}
  {\bibfnamefont {M.}~\bibnamefont {Grioni}}, \bibinfo {author} {\bibfnamefont
  {J.~V.}\ \bibnamefont {Lauritsen}}, \bibinfo {author} {\bibfnamefont {P.~D.}\
  \bibnamefont {King}}, \bibinfo {author} {\bibfnamefont {P.}~\bibnamefont
  {Hofmann}}, \ and\ \bibinfo {author} {\bibfnamefont {S.}~\bibnamefont
  {Ulstrup}},\ }\bibfield  {title} {\enquote {\bibinfo {title} {{Observation of
  Ultrafast Free Carrier Dynamics in Single Layer MoS2}},}\ }\href {\doibase
  10.1021/ACS.NANOLETT.5B01967/ASSET/IMAGES/LARGE/NL-2015-01967V_0005.JPEG}
  {\bibfield  {journal} {\bibinfo  {journal} {Nano Letters}\ }\textbf {\bibinfo
  {volume} {15}},\ \bibinfo {pages} {5883--5887} (\bibinfo {year}
  {2015})}\BibitemShut {NoStop}%
\bibitem [{\citenamefont {Klots}\ \emph {et~al.}(2014)\citenamefont {Klots},
  \citenamefont {Newaz}, \citenamefont {Wang}, \citenamefont {Prasai},
  \citenamefont {Krzyzanowska}, \citenamefont {Lin}, \citenamefont {Caudel},
  \citenamefont {Ghimire}, \citenamefont {Yan}, \citenamefont {Ivanov},
  \citenamefont {Velizhanin}, \citenamefont {Burger}, \citenamefont {Mandrus},
  \citenamefont {Tolk}, \citenamefont {Pantelides},\ and\ \citenamefont
  {Bolotin}}]{Klots2014}%
  \BibitemOpen
  \bibfield  {author} {\bibinfo {author} {\bibfnamefont {A.~R.}\ \bibnamefont
  {Klots}}, \bibinfo {author} {\bibfnamefont {A.~K.}\ \bibnamefont {Newaz}},
  \bibinfo {author} {\bibfnamefont {B.}~\bibnamefont {Wang}}, \bibinfo {author}
  {\bibfnamefont {D.}~\bibnamefont {Prasai}}, \bibinfo {author} {\bibfnamefont
  {H.}~\bibnamefont {Krzyzanowska}}, \bibinfo {author} {\bibfnamefont
  {J.}~\bibnamefont {Lin}}, \bibinfo {author} {\bibfnamefont {D.}~\bibnamefont
  {Caudel}}, \bibinfo {author} {\bibfnamefont {N.~J.}\ \bibnamefont {Ghimire}},
  \bibinfo {author} {\bibfnamefont {J.}~\bibnamefont {Yan}}, \bibinfo {author}
  {\bibfnamefont {B.~L.}\ \bibnamefont {Ivanov}}, \bibinfo {author}
  {\bibfnamefont {K.~A.}\ \bibnamefont {Velizhanin}}, \bibinfo {author}
  {\bibfnamefont {A.}~\bibnamefont {Burger}}, \bibinfo {author} {\bibfnamefont
  {D.~G.}\ \bibnamefont {Mandrus}}, \bibinfo {author} {\bibfnamefont {N.~H.}\
  \bibnamefont {Tolk}}, \bibinfo {author} {\bibfnamefont {S.~T.}\ \bibnamefont
  {Pantelides}}, \ and\ \bibinfo {author} {\bibfnamefont {K.~I.}\ \bibnamefont
  {Bolotin}},\ }\bibfield  {title} {\enquote {\bibinfo {title} {{Probing
  excitonic states in suspended two-dimensional semiconductors by photocurrent
  spectroscopy}},}\ }\href {\doibase 10.1038/srep06608} {\bibfield  {journal}
  {\bibinfo  {journal} {Scientific Reports}\ }\textbf {\bibinfo {volume} {4}},\
  \bibinfo {pages} {1--7} (\bibinfo {year} {2014})}\BibitemShut {NoStop}%
\bibitem [{\citenamefont {Qiu}, \citenamefont {{Da Jornada}},\ and\
  \citenamefont {Louie}(2013)}]{Qiu2013}%
  \BibitemOpen
  \bibfield  {author} {\bibinfo {author} {\bibfnamefont {D.~Y.}\ \bibnamefont
  {Qiu}}, \bibinfo {author} {\bibfnamefont {F.~H.}\ \bibnamefont {{Da
  Jornada}}}, \ and\ \bibinfo {author} {\bibfnamefont {S.~G.}\ \bibnamefont
  {Louie}},\ }\bibfield  {title} {\enquote {\bibinfo {title} {{Optical spectrum
  of MoS2: Many-body effects and diversity of exciton states}},}\ }\href
  {\doibase 10.1103/PHYSREVLETT.111.216805/FIGURES/3/MEDIUM} {\bibfield
  {journal} {\bibinfo  {journal} {Physical Review Letters}\ }\textbf {\bibinfo
  {volume} {111}},\ \bibinfo {pages} {216805} (\bibinfo {year}
  {2013})}\BibitemShut {NoStop}%
\bibitem [{\citenamefont {Molina-S{\'{a}}nchez}\ \emph
  {et~al.}(2013{\natexlab{a}})\citenamefont {Molina-S{\'{a}}nchez},
  \citenamefont {Sangalli}, \citenamefont {Hummer}, \citenamefont {Marini},\
  and\ \citenamefont {Wirtz}}]{Molina-Sanchez2013}%
  \BibitemOpen
  \bibfield  {author} {\bibinfo {author} {\bibfnamefont {A.}~\bibnamefont
  {Molina-S{\'{a}}nchez}}, \bibinfo {author} {\bibfnamefont {D.}~\bibnamefont
  {Sangalli}}, \bibinfo {author} {\bibfnamefont {K.}~\bibnamefont {Hummer}},
  \bibinfo {author} {\bibfnamefont {A.}~\bibnamefont {Marini}}, \ and\ \bibinfo
  {author} {\bibfnamefont {L.}~\bibnamefont {Wirtz}},\ }\bibfield  {title}
  {\enquote {\bibinfo {title} {{Effect of spin-orbit interaction on the optical
  spectra of single-layer, double-layer, and bulk MoS2}},}\ }\href {\doibase
  10.1103/PHYSREVB.88.045412/FIGURES/4/MEDIUM} {\bibfield  {journal} {\bibinfo
  {journal} {Physical Review B}\ }\textbf {\bibinfo {volume} {88}},\ \bibinfo
  {pages} {045412} (\bibinfo {year} {2013}{\natexlab{a}})}\BibitemShut
  {NoStop}%
\bibitem [{\citenamefont {Marcus}(1993)}]{Marcus1993}%
  \BibitemOpen
  \bibfield  {author} {\bibinfo {author} {\bibfnamefont {R.~A.}\ \bibnamefont
  {Marcus}},\ }\bibfield  {title} {\enquote {\bibinfo {title} {{Electron
  transfer reactions in chemistry. Theory and experiment}},}\ }\href {\doibase
  10.1103/RevModPhys.65.599} {\bibfield  {journal} {\bibinfo  {journal}
  {Reviews of Modern Physics}\ }\textbf {\bibinfo {volume} {65}},\ \bibinfo
  {pages} {599} (\bibinfo {year} {1993})}\BibitemShut {NoStop}%
\bibitem [{\citenamefont {Chidsey}(1991)}]{Chidsey1991a}%
  \BibitemOpen
  \bibfield  {author} {\bibinfo {author} {\bibfnamefont {C.~E.}\ \bibnamefont
  {Chidsey}},\ }\bibfield  {title} {\enquote {\bibinfo {title} {{Free Energy
  and Temperature Dependence of Electron Transfer at the Metal-Electrolyte
  Interface}},}\ }\href {\doibase 10.1126/SCIENCE.251.4996.919} {\bibfield
  {journal} {\bibinfo  {journal} {Science}\ }\textbf {\bibinfo {volume}
  {4996}},\ \bibinfo {pages} {919--922} (\bibinfo {year} {1991})}\BibitemShut
  {NoStop}%
\bibitem [{Note1()}]{Note1}%
  \BibitemOpen
  \bibinfo {note} {Marcus-type arguments have been previously employed to
  analyze charge transfer characteristics between TMDCs and quantum dots or
  molecular species.}\BibitemShut {Stop}%
\bibitem [{\citenamefont {Fang}\ \emph {et~al.}(2015)\citenamefont {Fang},
  \citenamefont {{Kuate Defo}}, \citenamefont {Shirodkar}, \citenamefont
  {Lieu}, \citenamefont {Tritsaris},\ and\ \citenamefont {Kaxiras}}]{Fang2015}%
  \BibitemOpen
  \bibfield  {author} {\bibinfo {author} {\bibfnamefont {S.}~\bibnamefont
  {Fang}}, \bibinfo {author} {\bibfnamefont {R.}~\bibnamefont {{Kuate Defo}}},
  \bibinfo {author} {\bibfnamefont {S.~N.}\ \bibnamefont {Shirodkar}}, \bibinfo
  {author} {\bibfnamefont {S.}~\bibnamefont {Lieu}}, \bibinfo {author}
  {\bibfnamefont {G.~A.}\ \bibnamefont {Tritsaris}}, \ and\ \bibinfo {author}
  {\bibfnamefont {E.}~\bibnamefont {Kaxiras}},\ }\bibfield  {title} {\enquote
  {\bibinfo {title} {{Ab initio tight-binding Hamiltonian for transition metal
  dichalcogenides}},}\ }\href {\doibase
  10.1103/PHYSREVB.92.205108/FIGURES/10/MEDIUM} {\bibfield  {journal} {\bibinfo
   {journal} {Physical Review B}\ }\textbf {\bibinfo {volume} {92}},\ \bibinfo
  {pages} {205108} (\bibinfo {year} {2015})}\BibitemShut {NoStop}%
\bibitem [{\citenamefont {Kadantsev}\ and\ \citenamefont
  {Hawrylak}(2012)}]{Kadantsev2012}%
  \BibitemOpen
  \bibfield  {author} {\bibinfo {author} {\bibfnamefont {E.~S.}\ \bibnamefont
  {Kadantsev}}\ and\ \bibinfo {author} {\bibfnamefont {P.}~\bibnamefont
  {Hawrylak}},\ }\bibfield  {title} {\enquote {\bibinfo {title} {{Electronic
  structure of a single MoS2 monolayer}},}\ }\href {\doibase
  10.1016/J.SSC.2012.02.005} {\bibfield  {journal} {\bibinfo  {journal} {Solid
  State Communications}\ }\textbf {\bibinfo {volume} {152}},\ \bibinfo {pages}
  {909--913} (\bibinfo {year} {2012})}\BibitemShut {NoStop}%
\bibitem [{\citenamefont {Korm{\'{a}}nyos}\ \emph {et~al.}(2015)\citenamefont
  {Korm{\'{a}}nyos}, \citenamefont {Burkard}, \citenamefont {Gmitra},
  \citenamefont {Fabian}, \citenamefont {Z{\'{o}}lyomi}, \citenamefont
  {Drummond},\ and\ \citenamefont {Fal'ko}}]{Kormanyos2015}%
  \BibitemOpen
  \bibfield  {author} {\bibinfo {author} {\bibfnamefont {A.}~\bibnamefont
  {Korm{\'{a}}nyos}}, \bibinfo {author} {\bibfnamefont {G.}~\bibnamefont
  {Burkard}}, \bibinfo {author} {\bibfnamefont {M.}~\bibnamefont {Gmitra}},
  \bibinfo {author} {\bibfnamefont {J.}~\bibnamefont {Fabian}}, \bibinfo
  {author} {\bibfnamefont {V.}~\bibnamefont {Z{\'{o}}lyomi}}, \bibinfo {author}
  {\bibfnamefont {N.~D.}\ \bibnamefont {Drummond}}, \ and\ \bibinfo {author}
  {\bibfnamefont {V.}~\bibnamefont {Fal'ko}},\ }\bibfield  {title} {\enquote
  {\bibinfo {title} {k{\textperiodcentered}p theory for two-dimensional
  transition metal dichalcogenide semiconductors},}\ }\href {\doibase
  10.1088/2053-1583/2/2/022001} {\bibfield  {journal} {\bibinfo  {journal} {2D
  Materials}\ }\textbf {\bibinfo {volume} {2}},\ \bibinfo {pages} {022001}
  (\bibinfo {year} {2015})}\BibitemShut {NoStop}%
\bibitem [{\citenamefont {Migliore}\ and\ \citenamefont
  {Nitzan}(2012)}]{Migliore2012}%
  \BibitemOpen
  \bibfield  {author} {\bibinfo {author} {\bibfnamefont {A.}~\bibnamefont
  {Migliore}}\ and\ \bibinfo {author} {\bibfnamefont {A.}~\bibnamefont
  {Nitzan}},\ }\bibfield  {title} {\enquote {\bibinfo {title} {{On the
  evaluation of the Marcus–Hush–Chidsey integral}},}\ }\href {\doibase
  10.1016/J.JELECHEM.2012.02.026} {\bibfield  {journal} {\bibinfo  {journal}
  {Journal of Electroanalytical Chemistry}\ }\textbf {\bibinfo {volume}
  {671}},\ \bibinfo {pages} {99--101} (\bibinfo {year} {2012})}\BibitemShut
  {NoStop}%
\bibitem [{\citenamefont {Ghosh}, \citenamefont {Soudackov},\ and\
  \citenamefont {Hammes-Schiffer}(2016)}]{Ghosh2016c}%
  \BibitemOpen
  \bibfield  {author} {\bibinfo {author} {\bibfnamefont {S.}~\bibnamefont
  {Ghosh}}, \bibinfo {author} {\bibfnamefont {A.~V.}\ \bibnamefont
  {Soudackov}}, \ and\ \bibinfo {author} {\bibfnamefont {S.}~\bibnamefont
  {Hammes-Schiffer}},\ }\bibfield  {title} {\enquote {\bibinfo {title}
  {{Electrochemical Electron Transfer and Proton-Coupled Electron Transfer:
  Effects of Double Layer and Ionic Environment on Solvent Reorganization
  Energies}},}\ }\href {\doibase
  10.1021/ACS.JCTC.6B00233/ASSET/IMAGES/LARGE/CT-2016-00233X_0010.JPEG}
  {\bibfield  {journal} {\bibinfo  {journal} {Journal of Chemical Theory and
  Computation}\ }\textbf {\bibinfo {volume} {12}},\ \bibinfo {pages}
  {2917--2925} (\bibinfo {year} {2016})}\BibitemShut {NoStop}%
\bibitem [{\citenamefont {Giustino}(2017)}]{Giustino2017}%
  \BibitemOpen
  \bibfield  {author} {\bibinfo {author} {\bibfnamefont {F.}~\bibnamefont
  {Giustino}},\ }\bibfield  {title} {\enquote {\bibinfo {title}
  {{Electron-phonon interactions from first principles}},}\ }\href {\doibase
  10.1103/REVMODPHYS.89.015003/FIGURES/18/MEDIUM} {\bibfield  {journal}
  {\bibinfo  {journal} {Reviews of Modern Physics}\ }\textbf {\bibinfo {volume}
  {89}},\ \bibinfo {pages} {015003} (\bibinfo {year} {2017})}\BibitemShut
  {NoStop}%
\bibitem [{\citenamefont {Zhou}\ \emph {et~al.}(2021)\citenamefont {Zhou},
  \citenamefont {Park}, \citenamefont {Timrov}, \citenamefont {Floris},
  \citenamefont {Cococcioni}, \citenamefont {Marzari},\ and\ \citenamefont
  {Bernardi}}]{Zhou2021}%
  \BibitemOpen
  \bibfield  {author} {\bibinfo {author} {\bibfnamefont {J.~J.}\ \bibnamefont
  {Zhou}}, \bibinfo {author} {\bibfnamefont {J.}~\bibnamefont {Park}}, \bibinfo
  {author} {\bibfnamefont {I.}~\bibnamefont {Timrov}}, \bibinfo {author}
  {\bibfnamefont {A.}~\bibnamefont {Floris}}, \bibinfo {author} {\bibfnamefont
  {M.}~\bibnamefont {Cococcioni}}, \bibinfo {author} {\bibfnamefont
  {N.}~\bibnamefont {Marzari}}, \ and\ \bibinfo {author} {\bibfnamefont
  {M.}~\bibnamefont {Bernardi}},\ }\bibfield  {title} {\enquote {\bibinfo
  {title} {{Ab Initio Electron-Phonon Interactions in Correlated Electron
  Systems}},}\ }\href {\doibase
  10.1103/PHYSREVLETT.127.126404/FIGURES/4/MEDIUM} {\bibfield  {journal}
  {\bibinfo  {journal} {Physical Review Letters}\ }\textbf {\bibinfo {volume}
  {127}},\ \bibinfo {pages} {126404} (\bibinfo {year} {2021})}\BibitemShut
  {NoStop}%
\bibitem [{\citenamefont {Chen}, \citenamefont {Sangalli},\ and\ \citenamefont
  {Bernardi}(2020)}]{Chen2020d}%
  \BibitemOpen
  \bibfield  {author} {\bibinfo {author} {\bibfnamefont {H.~Y.}\ \bibnamefont
  {Chen}}, \bibinfo {author} {\bibfnamefont {D.}~\bibnamefont {Sangalli}}, \
  and\ \bibinfo {author} {\bibfnamefont {M.}~\bibnamefont {Bernardi}},\
  }\bibfield  {title} {\enquote {\bibinfo {title} {{Exciton-Phonon Interaction
  and Relaxation Times from First Principles}},}\ }\href {\doibase
  10.1103/PHYSREVLETT.125.107401/FIGURES/5/MEDIUM} {\bibfield  {journal}
  {\bibinfo  {journal} {Physical Review Letters}\ }\textbf {\bibinfo {volume}
  {125}},\ \bibinfo {pages} {107401} (\bibinfo {year} {2020})}\BibitemShut
  {NoStop}%
\bibitem [{\citenamefont {Cunningham}\ \emph {et~al.}(2017)\citenamefont
  {Cunningham}, \citenamefont {Hanbicki}, \citenamefont {McCreary},\ and\
  \citenamefont {Jonker}}]{Cunningham2017}%
  \BibitemOpen
  \bibfield  {author} {\bibinfo {author} {\bibfnamefont {P.~D.}\ \bibnamefont
  {Cunningham}}, \bibinfo {author} {\bibfnamefont {A.~T.}\ \bibnamefont
  {Hanbicki}}, \bibinfo {author} {\bibfnamefont {K.~M.}\ \bibnamefont
  {McCreary}}, \ and\ \bibinfo {author} {\bibfnamefont {B.~T.}\ \bibnamefont
  {Jonker}},\ }\bibfield  {title} {\enquote {\bibinfo {title} {{Photoinduced
  Bandgap Renormalization and Exciton Binding Energy Reduction in WS2}},}\
  }\href {\doibase
  10.1021/ACSNANO.7B06885/ASSET/IMAGES/LARGE/NN-2017-06885E_0009.JPEG}
  {\bibfield  {journal} {\bibinfo  {journal} {ACS Nano}\ }\textbf {\bibinfo
  {volume} {11}},\ \bibinfo {pages} {12601--12608} (\bibinfo {year}
  {2017})}\BibitemShut {NoStop}%
\bibitem [{\citenamefont {Trovatello}\ \emph {et~al.}(2020)\citenamefont
  {Trovatello}, \citenamefont {Katsch}, \citenamefont {Borys}, \citenamefont
  {Selig}, \citenamefont {Yao}, \citenamefont {Borrego-Varillas}, \citenamefont
  {Scotognella}, \citenamefont {Kriegel}, \citenamefont {Yan}, \citenamefont
  {Zettl}, \citenamefont {Schuck}, \citenamefont {Knorr}, \citenamefont
  {Cerullo},\ and\ \citenamefont {Conte}}]{Trovatello2020}%
  \BibitemOpen
  \bibfield  {author} {\bibinfo {author} {\bibfnamefont {C.}~\bibnamefont
  {Trovatello}}, \bibinfo {author} {\bibfnamefont {F.}~\bibnamefont {Katsch}},
  \bibinfo {author} {\bibfnamefont {N.~J.}\ \bibnamefont {Borys}}, \bibinfo
  {author} {\bibfnamefont {M.}~\bibnamefont {Selig}}, \bibinfo {author}
  {\bibfnamefont {K.}~\bibnamefont {Yao}}, \bibinfo {author} {\bibfnamefont
  {R.}~\bibnamefont {Borrego-Varillas}}, \bibinfo {author} {\bibfnamefont
  {F.}~\bibnamefont {Scotognella}}, \bibinfo {author} {\bibfnamefont
  {I.}~\bibnamefont {Kriegel}}, \bibinfo {author} {\bibfnamefont
  {A.}~\bibnamefont {Yan}}, \bibinfo {author} {\bibfnamefont {A.}~\bibnamefont
  {Zettl}}, \bibinfo {author} {\bibfnamefont {P.~J.}\ \bibnamefont {Schuck}},
  \bibinfo {author} {\bibfnamefont {A.}~\bibnamefont {Knorr}}, \bibinfo
  {author} {\bibfnamefont {G.}~\bibnamefont {Cerullo}}, \ and\ \bibinfo
  {author} {\bibfnamefont {S.~D.}\ \bibnamefont {Conte}},\ }\bibfield  {title}
  {\enquote {\bibinfo {title} {{The ultrafast onset of exciton formation in 2D
  semiconductors}},}\ }\href {\doibase 10.1038/s41467-020-18835-5} {\bibfield
  {journal} {\bibinfo  {journal} {Nature Communications}\ }\textbf {\bibinfo
  {volume} {11}},\ \bibinfo {pages} {1--8} (\bibinfo {year}
  {2020})}\BibitemShut {NoStop}%
\bibitem [{\citenamefont {Aleithan}\ \emph {et~al.}(2016)\citenamefont
  {Aleithan}, \citenamefont {Livshits}, \citenamefont {Khadka}, \citenamefont
  {Rack}, \citenamefont {Kordesch},\ and\ \citenamefont
  {Stinaff}}]{Aleithan2016}%
  \BibitemOpen
  \bibfield  {author} {\bibinfo {author} {\bibfnamefont {S.~H.}\ \bibnamefont
  {Aleithan}}, \bibinfo {author} {\bibfnamefont {M.~Y.}\ \bibnamefont
  {Livshits}}, \bibinfo {author} {\bibfnamefont {S.}~\bibnamefont {Khadka}},
  \bibinfo {author} {\bibfnamefont {J.~J.}\ \bibnamefont {Rack}}, \bibinfo
  {author} {\bibfnamefont {M.~E.}\ \bibnamefont {Kordesch}}, \ and\ \bibinfo
  {author} {\bibfnamefont {E.}~\bibnamefont {Stinaff}},\ }\bibfield  {title}
  {\enquote {\bibinfo {title} {{Broadband femtosecond transient absorption
  spectroscopy for a CVD Mo S2 monolayer}},}\ }\href {\doibase
  10.1103/PHYSREVB.94.035445/FIGURES/8/MEDIUM} {\bibfield  {journal} {\bibinfo
  {journal} {Physical Review B}\ }\textbf {\bibinfo {volume} {94}},\ \bibinfo
  {pages} {035445} (\bibinfo {year} {2016})}\BibitemShut {NoStop}%
\bibitem [{\citenamefont {Gao}\ \emph {et~al.}(2021)\citenamefont {Gao},
  \citenamefont {Hu}, \citenamefont {Lu}, \citenamefont {Liu},\ and\
  \citenamefont {Ni}}]{Gao2021}%
  \BibitemOpen
  \bibfield  {author} {\bibinfo {author} {\bibfnamefont {L.}~\bibnamefont
  {Gao}}, \bibinfo {author} {\bibfnamefont {Z.}~\bibnamefont {Hu}}, \bibinfo
  {author} {\bibfnamefont {J.}~\bibnamefont {Lu}}, \bibinfo {author}
  {\bibfnamefont {H.}~\bibnamefont {Liu}}, \ and\ \bibinfo {author}
  {\bibfnamefont {Z.}~\bibnamefont {Ni}},\ }\bibfield  {title} {\enquote
  {\bibinfo {title} {{Defect-related dynamics of photoexcited carriers in 2D
  transition metal dichalcogenides}},}\ }\href {\doibase 10.1039/D1CP00006C}
  {\bibfield  {journal} {\bibinfo  {journal} {Physical Chemistry Chemical
  Physics}\ }\textbf {\bibinfo {volume} {23}},\ \bibinfo {pages} {8222--8235}
  (\bibinfo {year} {2021})}\BibitemShut {NoStop}%
\bibitem [{\citenamefont {Efimkin}\ \emph {et~al.}(2021)\citenamefont
  {Efimkin}, \citenamefont {Laird}, \citenamefont {Levinsen}, \citenamefont
  {Parish},\ and\ \citenamefont {Macdonald}}]{Efimkin2021}%
  \BibitemOpen
  \bibfield  {author} {\bibinfo {author} {\bibfnamefont {D.~K.}\ \bibnamefont
  {Efimkin}}, \bibinfo {author} {\bibfnamefont {E.~K.}\ \bibnamefont {Laird}},
  \bibinfo {author} {\bibfnamefont {J.}~\bibnamefont {Levinsen}}, \bibinfo
  {author} {\bibfnamefont {M.~M.}\ \bibnamefont {Parish}}, \ and\ \bibinfo
  {author} {\bibfnamefont {A.~H.}\ \bibnamefont {Macdonald}},\ }\bibfield
  {title} {\enquote {\bibinfo {title} {{Electron-exciton interactions in the
  exciton-polaron problem}},}\ }\href {\doibase
  10.1103/PHYSREVB.103.075417/FIGURES/10/MEDIUM} {\bibfield  {journal}
  {\bibinfo  {journal} {Physical Review B}\ }\textbf {\bibinfo {volume}
  {103}},\ \bibinfo {pages} {075417} (\bibinfo {year} {2021})}\BibitemShut
  {NoStop}%
\bibitem [{\citenamefont {Ataei}\ and\ \citenamefont
  {Sadeghi}(2021)}]{Ataei2021}%
  \BibitemOpen
  \bibfield  {author} {\bibinfo {author} {\bibfnamefont {S.~S.}\ \bibnamefont
  {Ataei}}\ and\ \bibinfo {author} {\bibfnamefont {A.}~\bibnamefont
  {Sadeghi}},\ }\bibfield  {title} {\enquote {\bibinfo {title} {{Competitive
  screening and band gap renormalization in n-type monolayer transition metal
  dichalcogenides}},}\ }\href {\doibase 10.1103/PhysRevB.104.155301} {\bibfield
   {journal} {\bibinfo  {journal} {Physical Review B}\ }\textbf {\bibinfo
  {volume} {104}},\ \bibinfo {pages} {155301} (\bibinfo {year}
  {2021})}\BibitemShut {NoStop}%
\bibitem [{\citenamefont {Chang}\ and\ \citenamefont
  {Reichman}(2019)}]{Chang2019b}%
  \BibitemOpen
  \bibfield  {author} {\bibinfo {author} {\bibfnamefont {Y.~W.}\ \bibnamefont
  {Chang}}\ and\ \bibinfo {author} {\bibfnamefont {D.~R.}\ \bibnamefont
  {Reichman}},\ }\bibfield  {title} {\enquote {\bibinfo {title} {{Many-body
  theory of optical absorption in doped two-dimensional semiconductors}},}\
  }\href {\doibase 10.1103/PHYSREVB.99.125421/FIGURES/4/MEDIUM} {\bibfield
  {journal} {\bibinfo  {journal} {Physical Review B}\ }\textbf {\bibinfo
  {volume} {99}},\ \bibinfo {pages} {125421} (\bibinfo {year}
  {2019})}\BibitemShut {NoStop}%
\bibitem [{\citenamefont {Cho}, \citenamefont {Bintrim},\ and\ \citenamefont
  {Berkelbach}(2022)}]{Cho2022}%
  \BibitemOpen
  \bibfield  {author} {\bibinfo {author} {\bibfnamefont {Y.}~\bibnamefont
  {Cho}}, \bibinfo {author} {\bibfnamefont {S.~J.}\ \bibnamefont {Bintrim}}, \
  and\ \bibinfo {author} {\bibfnamefont {T.~C.}\ \bibnamefont {Berkelbach}},\
  }\bibfield  {title} {\enquote {\bibinfo {title} {{Simplified GW/BSE Approach
  for Charged and Neutral Excitation Energies of Large Molecules and
  Nanomaterials}},}\ }\href {\doibase
  10.1021/ACS.JCTC.2C00087/ASSET/IMAGES/LARGE/CT2C00087_0005.JPEG} {\bibfield
  {journal} {\bibinfo  {journal} {Journal of Chemical Theory and Computation}\
  }\textbf {\bibinfo {volume} {18}},\ \bibinfo {pages} {3438--3446} (\bibinfo
  {year} {2022})}\BibitemShut {NoStop}%
\bibitem [{\citenamefont {Antonius}\ and\ \citenamefont
  {Louie}(2022)}]{Antonius2022}%
  \BibitemOpen
  \bibfield  {author} {\bibinfo {author} {\bibfnamefont {G.}~\bibnamefont
  {Antonius}}\ and\ \bibinfo {author} {\bibfnamefont {S.~G.}\ \bibnamefont
  {Louie}},\ }\bibfield  {title} {\enquote {\bibinfo {title} {{Theory of
  exciton-phonon coupling}},}\ }\href {\doibase
  10.1103/PHYSREVB.105.085111/FIGURES/10/MEDIUM} {\bibfield  {journal}
  {\bibinfo  {journal} {Physical Review B}\ }\textbf {\bibinfo {volume}
  {105}},\ \bibinfo {pages} {085111} (\bibinfo {year} {2022})}\BibitemShut
  {NoStop}%
\bibitem [{\citenamefont {Islam}\ \emph {et~al.}(2022)\citenamefont {Islam},
  \citenamefont {Ismael}, \citenamefont {Luthy}, \citenamefont {Kizilkaya},\
  and\ \citenamefont {Escarra}}]{Islam2022}%
  \BibitemOpen
  \bibfield  {author} {\bibinfo {author} {\bibfnamefont {K.~M.}\ \bibnamefont
  {Islam}}, \bibinfo {author} {\bibfnamefont {T.}~\bibnamefont {Ismael}},
  \bibinfo {author} {\bibfnamefont {C.}~\bibnamefont {Luthy}}, \bibinfo
  {author} {\bibfnamefont {O.}~\bibnamefont {Kizilkaya}}, \ and\ \bibinfo
  {author} {\bibfnamefont {M.~D.}\ \bibnamefont {Escarra}},\ }\bibfield
  {title} {\enquote {\bibinfo {title} {{Large-Area, High-Specific-Power
  Schottky-Junction Photovoltaics from CVD-Grown Monolayer MoS2}},}\ }\href
  {\doibase 10.1021/ACSAMI.2C01650/ASSET/IMAGES/LARGE/AM2C01650_0009.JPEG}
  {\bibfield  {journal} {\bibinfo  {journal} {Applied Materials and
  Interfaces}\ }\textbf {\bibinfo {volume} {14}},\ \bibinfo {pages}
  {24281--24289} (\bibinfo {year} {2022})}\BibitemShut {NoStop}%
\bibitem [{\citenamefont {Wang}\ \emph {et~al.}(2022)\citenamefont {Wang},
  \citenamefont {Kim}, \citenamefont {Li}, \citenamefont {Ma}, \citenamefont
  {Hong}, \citenamefont {Kim}, \citenamefont {Shin}, \citenamefont {Jeong},\
  and\ \citenamefont {Chhowalla}}]{Wang2022a}%
  \BibitemOpen
  \bibfield  {author} {\bibinfo {author} {\bibfnamefont {Y.}~\bibnamefont
  {Wang}}, \bibinfo {author} {\bibfnamefont {J.~C.}\ \bibnamefont {Kim}},
  \bibinfo {author} {\bibfnamefont {Y.}~\bibnamefont {Li}}, \bibinfo {author}
  {\bibfnamefont {K.~Y.}\ \bibnamefont {Ma}}, \bibinfo {author} {\bibfnamefont
  {S.}~\bibnamefont {Hong}}, \bibinfo {author} {\bibfnamefont {M.}~\bibnamefont
  {Kim}}, \bibinfo {author} {\bibfnamefont {H.~S.}\ \bibnamefont {Shin}},
  \bibinfo {author} {\bibfnamefont {H.~Y.}\ \bibnamefont {Jeong}}, \ and\
  \bibinfo {author} {\bibfnamefont {M.}~\bibnamefont {Chhowalla}},\ }\bibfield
  {title} {\enquote {\bibinfo {title} {{P-type electrical contacts for 2D
  transition-metal dichalcogenides}},}\ }\href {\doibase
  10.1038/s41586-022-05134-w} {\bibfield  {journal} {\bibinfo  {journal}
  {Nature}\ }\textbf {\bibinfo {volume} {610}},\ \bibinfo {pages} {61--66}
  (\bibinfo {year} {2022})}\BibitemShut {NoStop}%
\bibitem [{\citenamefont {Green}(2003)}]{Green2003}%
  \BibitemOpen
  \bibfield  {author} {\bibinfo {author} {\bibfnamefont {M.~A.}\ \bibnamefont
  {Green}},\ }\href {\doibase 10.1007/B137807} {\emph {\bibinfo {title} {{Third
  Generation Photovoltaics: Advanced Solar Energy Conversion}}}}\ (\bibinfo
  {publisher} {Springer Berlin Heidelberg},\ \bibinfo {year}
  {2003})\BibitemShut {NoStop}%
\bibitem [{\citenamefont {Wang}\ and\ \citenamefont
  {Sambur}(2019)}]{Wang2019f}%
  \BibitemOpen
  \bibfield  {author} {\bibinfo {author} {\bibfnamefont {L.}~\bibnamefont
  {Wang}}\ and\ \bibinfo {author} {\bibfnamefont {J.~B.}\ \bibnamefont
  {Sambur}},\ }\bibfield  {title} {\enquote {\bibinfo {title} {{Efficient
  Ultrathin Liquid Junction Photovoltaics Based on Transition Metal
  Dichalcogenides}},}\ }\href {\doibase
  10.1021/ACS.NANOLETT.9B00070/ASSET/IMAGES/LARGE/NL-2019-00070Z_0005.JPEG}
  {\bibfield  {journal} {\bibinfo  {journal} {Nano Letters}\ }\textbf {\bibinfo
  {volume} {19}},\ \bibinfo {pages} {2960--2967} (\bibinfo {year}
  {2019})}\BibitemShut {NoStop}%
\bibitem [{\citenamefont {Lee}\ \emph {et~al.}(2010)\citenamefont {Lee},
  \citenamefont {Yan}, \citenamefont {Brus}, \citenamefont {Heinz},
  \citenamefont {Hone},\ and\ \citenamefont {Ryu}}]{Lee2010}%
  \BibitemOpen
  \bibfield  {author} {\bibinfo {author} {\bibfnamefont {C.}~\bibnamefont
  {Lee}}, \bibinfo {author} {\bibfnamefont {H.}~\bibnamefont {Yan}}, \bibinfo
  {author} {\bibfnamefont {L.~E.}\ \bibnamefont {Brus}}, \bibinfo {author}
  {\bibfnamefont {T.~F.}\ \bibnamefont {Heinz}}, \bibinfo {author}
  {\bibfnamefont {J.}~\bibnamefont {Hone}}, \ and\ \bibinfo {author}
  {\bibfnamefont {S.}~\bibnamefont {Ryu}},\ }\bibfield  {title} {\enquote
  {\bibinfo {title} {{Anomalous lattice vibrations of single- and few-layer
  MoS2}},}\ }\href {\doibase 10.1021/NN1003937/SUPPL_FILE/NN1003937_SI_001.PDF}
  {\bibfield  {journal} {\bibinfo  {journal} {ACS Nano}\ }\textbf {\bibinfo
  {volume} {4}},\ \bibinfo {pages} {2695--2700} (\bibinfo {year}
  {2010})}\BibitemShut {NoStop}%
\bibitem [{\citenamefont {Lee}\ \emph {et~al.}(2014)\citenamefont {Lee},
  \citenamefont {Lee}, \citenamefont {{Van Der Zande}}, \citenamefont {Chen},
  \citenamefont {Li}, \citenamefont {Han}, \citenamefont {Cui}, \citenamefont
  {Arefe}, \citenamefont {Nuckolls}, \citenamefont {Heinz}, \citenamefont
  {Guo}, \citenamefont {Hone},\ and\ \citenamefont {Kim}}]{Lee2014a}%
  \BibitemOpen
  \bibfield  {author} {\bibinfo {author} {\bibfnamefont {C.~H.}\ \bibnamefont
  {Lee}}, \bibinfo {author} {\bibfnamefont {G.~H.}\ \bibnamefont {Lee}},
  \bibinfo {author} {\bibfnamefont {A.~M.}\ \bibnamefont {{Van Der Zande}}},
  \bibinfo {author} {\bibfnamefont {W.}~\bibnamefont {Chen}}, \bibinfo {author}
  {\bibfnamefont {Y.}~\bibnamefont {Li}}, \bibinfo {author} {\bibfnamefont
  {M.}~\bibnamefont {Han}}, \bibinfo {author} {\bibfnamefont {X.}~\bibnamefont
  {Cui}}, \bibinfo {author} {\bibfnamefont {G.}~\bibnamefont {Arefe}}, \bibinfo
  {author} {\bibfnamefont {C.}~\bibnamefont {Nuckolls}}, \bibinfo {author}
  {\bibfnamefont {T.~F.}\ \bibnamefont {Heinz}}, \bibinfo {author}
  {\bibfnamefont {J.}~\bibnamefont {Guo}}, \bibinfo {author} {\bibfnamefont
  {J.}~\bibnamefont {Hone}}, \ and\ \bibinfo {author} {\bibfnamefont
  {P.}~\bibnamefont {Kim}},\ }\bibfield  {title} {\enquote {\bibinfo {title}
  {{Atomically thin p–n junctions with van der Waals heterointerfaces}},}\
  }\href {\doibase 10.1038/nnano.2014.150} {\bibfield  {journal} {\bibinfo
  {journal} {Nature Nanotechnology}\ }\textbf {\bibinfo {volume} {9}},\
  \bibinfo {pages} {676--681} (\bibinfo {year} {2014})}\BibitemShut {NoStop}%
\bibitem [{\citenamefont {Baraff}\ and\ \citenamefont
  {Schiiter}(1984)}]{Baraff1984}%
  \BibitemOpen
  \bibfield  {author} {\bibinfo {author} {\bibfnamefont {G.~A.}\ \bibnamefont
  {Baraff}}\ and\ \bibinfo {author} {\bibfnamefont {M.}~\bibnamefont
  {Schiiter}},\ }\bibfield  {title} {\enquote {\bibinfo {title} {{Migration of
  interstitials in silicon}},}\ }\href {\doibase 10.1103/PhysRevB.30.3460}
  {\bibfield  {journal} {\bibinfo  {journal} {Physical Review B}\ }\textbf
  {\bibinfo {volume} {30}},\ \bibinfo {pages} {3460} (\bibinfo {year}
  {1984})}\BibitemShut {NoStop}%
\bibitem [{\citenamefont {Giannozzi}\ \emph {et~al.}(2017)\citenamefont
  {Giannozzi}, \citenamefont {Andreussi}, \citenamefont {Brumme}, \citenamefont
  {Bunau}, \citenamefont {{Buongiorno Nardelli}}, \citenamefont {Calandra},
  \citenamefont {Car}, \citenamefont {Cavazzoni}, \citenamefont {Ceresoli},
  \citenamefont {Cococcioni}, \citenamefont {Colonna}, \citenamefont
  {Carnimeo}, \citenamefont {{Dal Corso}}, \citenamefont {{De Gironcoli}},
  \citenamefont {Delugas}, \citenamefont {Distasio}, \citenamefont {Ferretti},
  \citenamefont {Floris}, \citenamefont {Fratesi}, \citenamefont {Fugallo},
  \citenamefont {Gebauer}, \citenamefont {Gerstmann}, \citenamefont {Giustino},
  \citenamefont {Gorni}, \citenamefont {Jia}, \citenamefont {Kawamura},
  \citenamefont {Ko}, \citenamefont {Kokalj}, \citenamefont
  {K{\"{u}}c{\"{u}}kbenli}, \citenamefont {Lazzeri}, \citenamefont {Marsili},
  \citenamefont {Marzari}, \citenamefont {Mauri}, \citenamefont {Nguyen},
  \citenamefont {Nguyen}, \citenamefont {Otero-De-La-Roza}, \citenamefont
  {Paulatto}, \citenamefont {Ponc{\'{e}}}, \citenamefont {Rocca}, \citenamefont
  {Sabatini}, \citenamefont {Santra}, \citenamefont {Schlipf}, \citenamefont
  {Seitsonen}, \citenamefont {Smogunov}, \citenamefont {Timrov}, \citenamefont
  {Thonhauser}, \citenamefont {Umari}, \citenamefont {Vast}, \citenamefont
  {Wu},\ and\ \citenamefont {Baroni}}]{Giannozzi2017}%
  \BibitemOpen
  \bibfield  {author} {\bibinfo {author} {\bibfnamefont {P.}~\bibnamefont
  {Giannozzi}}, \bibinfo {author} {\bibfnamefont {O.}~\bibnamefont
  {Andreussi}}, \bibinfo {author} {\bibfnamefont {T.}~\bibnamefont {Brumme}},
  \bibinfo {author} {\bibfnamefont {O.}~\bibnamefont {Bunau}}, \bibinfo
  {author} {\bibfnamefont {M.}~\bibnamefont {{Buongiorno Nardelli}}}, \bibinfo
  {author} {\bibfnamefont {M.}~\bibnamefont {Calandra}}, \bibinfo {author}
  {\bibfnamefont {R.}~\bibnamefont {Car}}, \bibinfo {author} {\bibfnamefont
  {C.}~\bibnamefont {Cavazzoni}}, \bibinfo {author} {\bibfnamefont
  {D.}~\bibnamefont {Ceresoli}}, \bibinfo {author} {\bibfnamefont
  {M.}~\bibnamefont {Cococcioni}}, \bibinfo {author} {\bibfnamefont
  {N.}~\bibnamefont {Colonna}}, \bibinfo {author} {\bibfnamefont
  {I.}~\bibnamefont {Carnimeo}}, \bibinfo {author} {\bibfnamefont
  {A.}~\bibnamefont {{Dal Corso}}}, \bibinfo {author} {\bibfnamefont
  {S.}~\bibnamefont {{De Gironcoli}}}, \bibinfo {author} {\bibfnamefont
  {P.}~\bibnamefont {Delugas}}, \bibinfo {author} {\bibfnamefont {R.~A.}\
  \bibnamefont {Distasio}}, \bibinfo {author} {\bibfnamefont {A.}~\bibnamefont
  {Ferretti}}, \bibinfo {author} {\bibfnamefont {A.}~\bibnamefont {Floris}},
  \bibinfo {author} {\bibfnamefont {G.}~\bibnamefont {Fratesi}}, \bibinfo
  {author} {\bibfnamefont {G.}~\bibnamefont {Fugallo}}, \bibinfo {author}
  {\bibfnamefont {R.}~\bibnamefont {Gebauer}}, \bibinfo {author} {\bibfnamefont
  {U.}~\bibnamefont {Gerstmann}}, \bibinfo {author} {\bibfnamefont
  {F.}~\bibnamefont {Giustino}}, \bibinfo {author} {\bibfnamefont
  {T.}~\bibnamefont {Gorni}}, \bibinfo {author} {\bibfnamefont
  {J.}~\bibnamefont {Jia}}, \bibinfo {author} {\bibfnamefont {M.}~\bibnamefont
  {Kawamura}}, \bibinfo {author} {\bibfnamefont {H.~Y.}\ \bibnamefont {Ko}},
  \bibinfo {author} {\bibfnamefont {A.}~\bibnamefont {Kokalj}}, \bibinfo
  {author} {\bibfnamefont {E.}~\bibnamefont {K{\"{u}}c{\"{u}}kbenli}}, \bibinfo
  {author} {\bibfnamefont {M.}~\bibnamefont {Lazzeri}}, \bibinfo {author}
  {\bibfnamefont {M.}~\bibnamefont {Marsili}}, \bibinfo {author} {\bibfnamefont
  {N.}~\bibnamefont {Marzari}}, \bibinfo {author} {\bibfnamefont
  {F.}~\bibnamefont {Mauri}}, \bibinfo {author} {\bibfnamefont {N.~L.}\
  \bibnamefont {Nguyen}}, \bibinfo {author} {\bibfnamefont {H.~V.}\
  \bibnamefont {Nguyen}}, \bibinfo {author} {\bibfnamefont {A.}~\bibnamefont
  {Otero-De-La-Roza}}, \bibinfo {author} {\bibfnamefont {L.}~\bibnamefont
  {Paulatto}}, \bibinfo {author} {\bibfnamefont {S.}~\bibnamefont
  {Ponc{\'{e}}}}, \bibinfo {author} {\bibfnamefont {D.}~\bibnamefont {Rocca}},
  \bibinfo {author} {\bibfnamefont {R.}~\bibnamefont {Sabatini}}, \bibinfo
  {author} {\bibfnamefont {B.}~\bibnamefont {Santra}}, \bibinfo {author}
  {\bibfnamefont {M.}~\bibnamefont {Schlipf}}, \bibinfo {author} {\bibfnamefont
  {A.~P.}\ \bibnamefont {Seitsonen}}, \bibinfo {author} {\bibfnamefont
  {A.}~\bibnamefont {Smogunov}}, \bibinfo {author} {\bibfnamefont
  {I.}~\bibnamefont {Timrov}}, \bibinfo {author} {\bibfnamefont
  {T.}~\bibnamefont {Thonhauser}}, \bibinfo {author} {\bibfnamefont
  {P.}~\bibnamefont {Umari}}, \bibinfo {author} {\bibfnamefont
  {N.}~\bibnamefont {Vast}}, \bibinfo {author} {\bibfnamefont {X.}~\bibnamefont
  {Wu}}, \ and\ \bibinfo {author} {\bibfnamefont {S.}~\bibnamefont {Baroni}},\
  }\bibfield  {title} {\enquote {\bibinfo {title} {{Advanced capabilities for
  materials modelling with Quantum ESPRESSO}},}\ }\href {\doibase
  10.1088/1361-648X/AA8F79} {\bibfield  {journal} {\bibinfo  {journal} {Journal
  of Physics: Condensed Matter}\ }\textbf {\bibinfo {volume} {29}},\ \bibinfo
  {pages} {465901} (\bibinfo {year} {2017})}\BibitemShut {NoStop}%
\bibitem [{\citenamefont {Sangalli}\ \emph {et~al.}(2019)\citenamefont
  {Sangalli}, \citenamefont {Ferretti}, \citenamefont {Miranda}, \citenamefont
  {Attaccalite}, \citenamefont {Marri}, \citenamefont {Cannuccia},
  \citenamefont {Melo}, \citenamefont {Marsili}, \citenamefont {Paleari},
  \citenamefont {Marrazzo}, \citenamefont {Prandini}, \citenamefont
  {Bonf{\`{a}}}, \citenamefont {Atambo}, \citenamefont {Affinito},
  \citenamefont {Palummo}, \citenamefont {Molina-S{\'{a}}nchez}, \citenamefont
  {Hogan}, \citenamefont {Gr{\"{u}}ning}, \citenamefont {Varsano},\ and\
  \citenamefont {Marini}}]{Sangalli2019}%
  \BibitemOpen
  \bibfield  {author} {\bibinfo {author} {\bibfnamefont {D.}~\bibnamefont
  {Sangalli}}, \bibinfo {author} {\bibfnamefont {A.}~\bibnamefont {Ferretti}},
  \bibinfo {author} {\bibfnamefont {H.}~\bibnamefont {Miranda}}, \bibinfo
  {author} {\bibfnamefont {C.}~\bibnamefont {Attaccalite}}, \bibinfo {author}
  {\bibfnamefont {I.}~\bibnamefont {Marri}}, \bibinfo {author} {\bibfnamefont
  {E.}~\bibnamefont {Cannuccia}}, \bibinfo {author} {\bibfnamefont
  {P.}~\bibnamefont {Melo}}, \bibinfo {author} {\bibfnamefont {M.}~\bibnamefont
  {Marsili}}, \bibinfo {author} {\bibfnamefont {F.}~\bibnamefont {Paleari}},
  \bibinfo {author} {\bibfnamefont {A.}~\bibnamefont {Marrazzo}}, \bibinfo
  {author} {\bibfnamefont {G.}~\bibnamefont {Prandini}}, \bibinfo {author}
  {\bibfnamefont {P.}~\bibnamefont {Bonf{\`{a}}}}, \bibinfo {author}
  {\bibfnamefont {M.~O.}\ \bibnamefont {Atambo}}, \bibinfo {author}
  {\bibfnamefont {F.}~\bibnamefont {Affinito}}, \bibinfo {author}
  {\bibfnamefont {M.}~\bibnamefont {Palummo}}, \bibinfo {author} {\bibfnamefont
  {A.}~\bibnamefont {Molina-S{\'{a}}nchez}}, \bibinfo {author} {\bibfnamefont
  {C.}~\bibnamefont {Hogan}}, \bibinfo {author} {\bibfnamefont
  {M.}~\bibnamefont {Gr{\"{u}}ning}}, \bibinfo {author} {\bibfnamefont
  {D.}~\bibnamefont {Varsano}}, \ and\ \bibinfo {author} {\bibfnamefont
  {A.}~\bibnamefont {Marini}},\ }\bibfield  {title} {\enquote {\bibinfo {title}
  {{Many-body perturbation theory calculations using the yambo code}},}\ }\href
  {\doibase 10.1088/1361-648X/AB15D0} {\bibfield  {journal} {\bibinfo
  {journal} {Journal of Physics: Condensed Matter}\ }\textbf {\bibinfo {volume}
  {31}},\ \bibinfo {pages} {325902} (\bibinfo {year} {2019})}\BibitemShut
  {NoStop}%
\bibitem [{\citenamefont {Molina-S{\'{a}}nchez}\ \emph
  {et~al.}(2013{\natexlab{b}})\citenamefont {Molina-S{\'{a}}nchez},
  \citenamefont {Sangalli}, \citenamefont {Hummer}, \citenamefont {Marini},\
  and\ \citenamefont {Wirtz}}]{Molina-Sanchez2013a}%
  \BibitemOpen
  \bibfield  {author} {\bibinfo {author} {\bibfnamefont {A.}~\bibnamefont
  {Molina-S{\'{a}}nchez}}, \bibinfo {author} {\bibfnamefont {D.}~\bibnamefont
  {Sangalli}}, \bibinfo {author} {\bibfnamefont {K.}~\bibnamefont {Hummer}},
  \bibinfo {author} {\bibfnamefont {A.}~\bibnamefont {Marini}}, \ and\ \bibinfo
  {author} {\bibfnamefont {L.}~\bibnamefont {Wirtz}},\ }\bibfield  {title}
  {\enquote {\bibinfo {title} {{Effect of spin-orbit interaction on the optical
  spectra of single-layer, double-layer, and bulk MoS2}},}\ }\href {\doibase
  10.1103/PHYSREVB.88.045412/FIGURES/4/MEDIUM} {\bibfield  {journal} {\bibinfo
  {journal} {Physical Review B}\ }\textbf {\bibinfo {volume} {88}},\ \bibinfo
  {pages} {045412} (\bibinfo {year} {2013}{\natexlab{b}})}\BibitemShut
  {NoStop}%
\bibitem [{\citenamefont {Qiu}, \citenamefont {{Da Jornada}},\ and\
  \citenamefont {Louie}(2015)}]{Qiu2015}%
  \BibitemOpen
  \bibfield  {author} {\bibinfo {author} {\bibfnamefont {D.~Y.}\ \bibnamefont
  {Qiu}}, \bibinfo {author} {\bibfnamefont {F.~H.}\ \bibnamefont {{Da
  Jornada}}}, \ and\ \bibinfo {author} {\bibfnamefont {S.~G.}\ \bibnamefont
  {Louie}},\ }\bibfield  {title} {\enquote {\bibinfo {title} {{Erratum: Optical
  Spectrum of MoS2: Many-Body Effects and Diversity of Exciton States (Physical
  Review Letters (2013) 111 (216805))}},}\ }\href {\doibase
  10.1103/PHYSREVLETT.115.119901/FIGURES/1/MEDIUM} {\bibfield  {journal}
  {\bibinfo  {journal} {Physical Review Letters}\ }\textbf {\bibinfo {volume}
  {115}},\ \bibinfo {pages} {119901} (\bibinfo {year} {2015})}\BibitemShut
  {NoStop}%
\bibitem [{\citenamefont {Berkelbach}\ and\ \citenamefont
  {Reichman}(2018)}]{Berkelbach2018b}%
  \BibitemOpen
  \bibfield  {author} {\bibinfo {author} {\bibfnamefont {T.~C.}\ \bibnamefont
  {Berkelbach}}\ and\ \bibinfo {author} {\bibfnamefont {D.~R.}\ \bibnamefont
  {Reichman}},\ }\bibfield  {title} {\enquote {\bibinfo {title} {{Optical and
  Excitonic Properties of Atomically Thin Transition-Metal Dichalcogenides}},}\
  }\href {\doibase 10.1146/ANNUREV-CONMATPHYS-033117-054009} {\bibfield
  {journal} {\bibinfo  {journal} {Annual Review of Condensed Matter Physics}\
  }\textbf {\bibinfo {volume} {9}},\ \bibinfo {pages} {379--396} (\bibinfo
  {year} {2018})}\BibitemShut {NoStop}%
\bibitem [{\citenamefont {Shi}\ \emph {et~al.}(2013)\citenamefont {Shi},
  \citenamefont {Yan}, \citenamefont {Bertolazzi}, \citenamefont {Brivio},
  \citenamefont {Gao}, \citenamefont {Kis}, \citenamefont {Jena}, \citenamefont
  {Xing},\ and\ \citenamefont {Huang}}]{Shi2013}%
  \BibitemOpen
  \bibfield  {author} {\bibinfo {author} {\bibfnamefont {H.}~\bibnamefont
  {Shi}}, \bibinfo {author} {\bibfnamefont {R.}~\bibnamefont {Yan}}, \bibinfo
  {author} {\bibfnamefont {S.}~\bibnamefont {Bertolazzi}}, \bibinfo {author}
  {\bibfnamefont {J.}~\bibnamefont {Brivio}}, \bibinfo {author} {\bibfnamefont
  {B.}~\bibnamefont {Gao}}, \bibinfo {author} {\bibfnamefont {A.}~\bibnamefont
  {Kis}}, \bibinfo {author} {\bibfnamefont {D.}~\bibnamefont {Jena}}, \bibinfo
  {author} {\bibfnamefont {H.~G.}\ \bibnamefont {Xing}}, \ and\ \bibinfo
  {author} {\bibfnamefont {L.}~\bibnamefont {Huang}},\ }\bibfield  {title}
  {\enquote {\bibinfo {title} {{Exciton dynamics in suspended monolayer and
  few-layer MoS2 2D crystals}},}\ }\href {\doibase
  10.1021/NN303973R/SUPPL_FILE/NN303973R_SI_001.PDF} {\bibfield  {journal}
  {\bibinfo  {journal} {ACS Nano}\ }\textbf {\bibinfo {volume} {7}},\ \bibinfo
  {pages} {1072--1080} (\bibinfo {year} {2013})}\BibitemShut {NoStop}%
\bibitem [{\citenamefont {{Dal Conte}}\ \emph {et~al.}(2020)\citenamefont {{Dal
  Conte}}, \citenamefont {Trovatello}, \citenamefont {Gadermaier},\ and\
  \citenamefont {Cerullo}}]{DalConte2020}%
  \BibitemOpen
  \bibfield  {author} {\bibinfo {author} {\bibfnamefont {S.}~\bibnamefont {{Dal
  Conte}}}, \bibinfo {author} {\bibfnamefont {C.}~\bibnamefont {Trovatello}},
  \bibinfo {author} {\bibfnamefont {C.}~\bibnamefont {Gadermaier}}, \ and\
  \bibinfo {author} {\bibfnamefont {G.}~\bibnamefont {Cerullo}},\ }\bibfield
  {title} {\enquote {\bibinfo {title} {{Ultrafast Photophysics of 2D
  Semiconductors and Related Heterostructures}},}\ }\href {\doibase
  10.1016/J.TRECHM.2019.07.007} {\bibfield  {journal} {\bibinfo  {journal}
  {Trends in Chemistry}\ }\textbf {\bibinfo {volume} {2}},\ \bibinfo {pages}
  {28--42} (\bibinfo {year} {2020})}\BibitemShut {NoStop}%
\bibitem [{\citenamefont {Ruckebusch}\ \emph {et~al.}(2012)\citenamefont
  {Ruckebusch}, \citenamefont {Sliwa}, \citenamefont {Pernot}, \citenamefont
  {de~Juan},\ and\ \citenamefont {Tauler}}]{Ruckebusch2012}%
  \BibitemOpen
  \bibfield  {author} {\bibinfo {author} {\bibfnamefont {C.}~\bibnamefont
  {Ruckebusch}}, \bibinfo {author} {\bibfnamefont {M.}~\bibnamefont {Sliwa}},
  \bibinfo {author} {\bibfnamefont {P.}~\bibnamefont {Pernot}}, \bibinfo
  {author} {\bibfnamefont {A.}~\bibnamefont {de~Juan}}, \ and\ \bibinfo
  {author} {\bibfnamefont {R.}~\bibnamefont {Tauler}},\ }\bibfield  {title}
  {\enquote {\bibinfo {title} {{Comprehensive data analysis of femtosecond
  transient absorption spectra: A review}},}\ }\href {\doibase
  10.1016/J.JPHOTOCHEMREV.2011.10.002} {\bibfield  {journal} {\bibinfo
  {journal} {Journal of Photochemistry and Photobiology C: Photochemistry
  Reviews}\ }\textbf {\bibinfo {volume} {13}},\ \bibinfo {pages} {1--27}
  (\bibinfo {year} {2012})}\BibitemShut {NoStop}%
\end{thebibliography}%

\end{document}